\documentclass[fleqn,usenatbib]{mnras}
\usepackage[utf8]{inputenc}

\usepackage[T1]{fontenc}

\DeclareRobustCommand{\VAN}[3]{#2}
\let\VANthebibliography\thebibliography
\def\thebibliography{\DeclareRobustCommand{\VAN}[3]{##3}\VANthebibliography}

\usepackage{graphicx}	% Including figure files
\usepackage{amsmath}	% Advanced maths commands
\usepackage{comment}
\usepackage{txfonts}
\usepackage{hyperref}
\usepackage{textgreek}
\usepackage{xargs}
\usepackage{xspace}
\usepackage{booktabs} %for BAGN table
\usepackage[dvipsnames,hyperref]{xcolor}

\usepackage{hyperref}
\hypersetup{
    colorlinks=true,
    linkcolor=Cerulean,
    citecolor=NavyBlue,
    filecolor=magenta,
    urlcolor=Violet,
    pdftitle= {LAP1-overview},
    pdfpagemode=FullScreen,
    }

\newcommand{\Msun}{\ensuremath{\mathrm{M}_\odot}\xspace}
\newcommand{\jwst}{\textit{JWST}\xspace}

\newcommand{\Halpha}{\text{H\textalpha}\xspace}
\newcommand{\Lya}{\text{Ly\textalpha}\xspace}
\newcommand{\Hbeta}{\text{H\textbeta}\xspace}
\newcommandx{\permittedEL}[6][1=O,2=III,3=,4=,5=,6=]{\text{{#1}\,{\sc {#2}}{#3}{#4}{#5}{#6}}\xspace}
\newcommandx{\semiforbiddenEL}[6][1=O,2=III,3=,4=,5=,6=]{\text{{#1}\,{\sc {#2}}]{#3}{#4}{#5}{#6}}\xspace}
\newcommandx{\forbiddenEL}[6][1=O,2=III,3=,4=,5=,6=]{\text{[{#1}\,{\sc{#2}}]{#3}{#4}{#5}{#6}}\xspace}
\newcommand{\kms}{km s$^{-1}$}
\newcommand{\HeIIL}[1][1=1640]{\permittedEL[He][ii][\textlambda][#1]}
\newcommandx{\OIIL}[1][1=3728]{\forbiddenEL[O][ii][\textlambda][#1]}
\newcommand{\OIIall}{\forbiddenEL[O][ii][\textlambda][\textlambda][3727,][30]}
\newcommandx{\OIIIL}[1][1=5008]{\forbiddenEL[O][iii][\textlambda][#1]}
\newcommand{\OIIIall}{\forbiddenEL[O][iii][\textlambda][\textlambda][5008,][4960]}
\newcommandx{\NIIL}[1][1=6585]{\forbiddenEL[N][ii][\textlambda][#1]}

\newcommandx{\NeIIIL}[1][1=3869]{\forbiddenEL[Ne][iii][\textlambda][#1]}

\newcommand{\CIIIall}{\semiforbiddenEL[C][iii][\textlambda][\textlambda][1907,][09]}
\newcommand{\CIVall}{\permittedEL[C][iv][\textlambda][\textlambda][1548,][51]}
\newcommand{\OIIIUvall}{\semiforbiddenEL[O][iii][\textlambda][\textlambda][1661,][66]}
\newcommand{\target}{LAP1\xspace}
\newcommand{\ergscm}{erg s$^{-1}$ cm$^{-2}\,$}
\newcommand{\JWST}{\textit{JWST}\xspace}

\newcommand{\msun}{\ensuremath{\mathrm{M_\odot}}\xspace}

\newcommand{\Mdyn}{\ensuremath{\mathrm{M_{\rm dyn}}}\xspace}

\newcommand{\fescLya}{$f_{\rm esc}^{\rm Ly\alpha}$}

\newcommand{\ksi}{$\xi_{\rm ion}$}
\newcommand{\ksio}{\xi_{\rm ion, 0}}

\newcommand{\prismifs}{PRISM-IFS\xspace}
\newcommand{\highresifs}{R2700-IFS\xspace}
\newcommand{\MSA}{R1000-MSA\xspace}

\title[LAP1]{GA-NIFS view of LAP1: Assembly of Pop~III stellar clusters during EoR}

\author[Jan Scholtz]{\parbox[h]{\textwidth}{
J. Scholtz$^{1,2}$,
R. Maiolino$^{1,2,3}$,
S. Carniani$^{4}$,
E. Vanzella$^{5}$,
R. J. Terlevich$^{6,7,8}$,
H. \"{U}bler$^{9}$, 
M. Perna$^{10}$,
S. Arribas$^{10}$,
E. Bertola$^{11}$,
P. Bergamini$^{5}$,
T. B\"oker$^{12}$,
A. Bolamperti$^{13,14}$,
A. Bunker$^{15}$,
S. Charlot$^{16}$,
M. Curti$^{5}$,
F. D'Eugenio$^{1,2}$,
C. Grillo$^{17, 18}$,
L. R. Ivey$^{1,2}$,
X. Ji$^{1,2}$,
G. C. Jones$^{1,2}$,
M. Meneghetti$^{5}$,
M. Messa$^{5}$,
R. Pascalau$^{1,2}$,
B. Rodríguez Del Pino$^{10}$,
P. Rosati$^{18,5}$,
E. Rusta$^{19,11}$,
S. Salvadori$^{19,11}$,
E. Terlevich$^{6,7,8}$,
G. Venturi$^{4}$,
S. Zamora$^{4}$,
A. Zanella$^{5}$
}\vspace{0.4cm}
\\
$^{1}$Kavli Institute for Cosmology, University of Cambridge, Madingley Road, Cambridge, CB3 0HA, UK\\
$^{2}$Cavendish Laboratory, University of Cambridge, 19 JJ Thomson Avenue, Cambridge CB3 0HE, UK\\
$^{3}$Department of Physics and Astronomy, University College London, Gower Street, London WC1E 6BT, UK\\
$^{4}$ Scuola Normale Superiore, Piazza dei Cavalieri 7, I-56126 Pisa, Italy\\
$^{5}$ INAF – OAS, Osservatorio di Astrofisica e Scienza dello Spazio di Bologna, via Gobetti 93/3, I-40129 Bologna, Italy\\
$^{6}$ Instituto Nacional de Astrofísica, Óptica y Electrónica, Tonantzintla, AP 51 y 216, 72000, Puebla, Mexico\\
$^{7}$ Institute of Astronomy, University of Cambridge, Cambridge, CB3 0HA, UK\\
$^{8}$ Facultad de Astronomía y Geofísica, Universidad de La Plata, FWA, B1900 La Plata, Argentina\\
$^{9}$ Max-Planck-Institut f\"ur extraterrestrische Physik, Gie{\ss}enbachstra{\ss}e 1, 85748 Garching, Germany\\
$^{10}$ Centro de Astrobiolog\'ia (CAB), CSIC--INTA, Cra. de Ajalvir Km.~4, 28850 -- Torrej\'on de Ardoz, Madrid, Spain \\
$^{11}$ INAF — Osservatorio Astrofisico di Arcetri, Largo E. Fermi 5, I-50125, Florence, Italy\\
$^{12}$ European Space Agency, c/o STScI, 3700 San Martin Drive, Baltimore MD 21218, USA\\
$^{13}$ Max-Planck-Institut f\"ur Astrophysik, Karl-Schwarzschild-Str. 1, D-85748 Garching, Germany \\
$^{14}$ INAF – IASF Milano, via A. Corti 12, I-20133 Milano, Italy\\
$^{15}$ Department of Physics, University of Oxford, Denys Wilkinson Building, Keble Road, Oxford OX1 3RH, UK \\
$^{16}$ Sorbonne Universit\'e, CNRS, UMR 7095, Institut d'Astrophysique de Paris, 98 bis bd Arago, 75014 Paris, France \\
$^{17}$ Dipartimento di Fisica, Universit\`a degli Studi di Milano, Via Celoria 16, I-20133 Milano, Italy\\ 
$^{18}$ Dipartimento di Fisica e Scienze della Terra, Università degli Studi di Ferrara, Via Saragat 1, I-44122 Ferrara, Italy\\
$^{19}$ Dipartimento di Fisica e Astronomia, Università degli Studi di Firenze, Largo E. Fermi 1, 50125, Firenze, Italy \\
}

\date{Accepted XXX. Received YYY; in original form ZZZ}

\pubyear{2026}

\begin{document}
\label{firstpage}
\pagerange{\pageref{firstpage}--\pageref{lastpage}}
\maketitle

\begin{abstract}
Lensed and Pristine 1 (LAP1) is one of the most promising candidates for the detection of so-called Pop~III stars at $z\sim6.6$. In this work, we present new high-spectral-resolution ($R \sim 2700$) \jwst NIRSpec-IFS observations of the system along with a detailed re-analysis of all archival \jwst and VLT/MUSE data. We confirm the very low metallicity of the system, with $12+\log(\text{O/H})=6.5\text{--}6.8$ (1--2\% Z$_{\odot}$), and its low stellar mass, with $M_\star= 10^{3\text{--}4}$ \Msun, based on the non-detection in \jwst/NIRCam filter stacks and the \Hbeta luminosity. The high-resolution IFS data allow us to constrain the upper limits on the dynamical masses ($\log_{10}(\Mdyn/\Msun)$) of the A and B components to $<6.6\text{--}7.6$ and $<5.6\text{--}6.6$, respectively. Based on the metallicity and the high equivalent width of \Halpha (EW $>1200$\,\AA), we estimate the possible range of the LAP1 stellar population ages to be $2\text{--}5$~Myr, consistent with the first burst of star formation in a pristine environment. Based on the size ($<10$~pc) and mass of the system, we discuss the nature of LAP1 as representing the formation of the first stellar clusters in a larger undetected galaxy, similar to other low-metallicity Pop~III candidates in the literature. 
\end{abstract}


%%%%%%%%%%%%%%%%%%%%%%%%%%%%%%%%%%%%%%%%%%%%%%%%%%

%%%%%%%%%%%%%%%%% BODY OF PAPER %%%%%%%%%%%%%%%%%%

\section{Introduction} 

The James Webb Space Telescope (\jwst) has enabled the search for and characterisation of the first population of stars, formed out of pristine gas. These stars, often called Population~III (Pop~III) stars, are one of the most sought-after discoveries in modern astrophysics. They are theorised to form in galactic embryos within the first few hundred million years after the Big Bang \citep[e.g.,][]{Abel_2002,Yoshida_2003,Klessen_2023}. Recent \jwst observations have revealed galaxies with extremely low metallicities, in the range $\sim 10^{-3}\text{--}10^{-2}~Z_\odot$ \citep[e.g.,][]{vanzella_extremely_2023, Fujimoto_2025, Hsiao2025_lowZ, maiolino_jades_2024-1, Morishita2025_lowZ, nakajima_ultra-faint_2025, Maiolino_QSO1_2026, Maiolino_2026, Uebler_2026,Vanzella_lap2_2026, Isobe_2026}. These objects could be very low-metallicity Pop~II galaxies, or possibly self-polluted or hybrid Pop~III galaxies \citep[][]{rusta_metal-polluted_2025}.

The main challenge in detecting Pop~III stars is the extremely short lifetime of their main spectroscopic features: 1) lack of metal lines; and 2) high equivalent width (EW) of He~II lines, both of which fade rapidly $\sim$3~Myr after the initial burst (e.g., \citealt{Schaerer_2002}), although models of rotating or binary Pop~III stars indicate longer lifetimes ($>10$~Myr) \citep[][]{Yoon2012, Lecroq2025}. While the lack of metal emission lines is due to the putative pristine composition of the interstellar medium (ISM) surrounding Pop~III stars, the strong He~II emission is directly linked to their metal-free composition. Stars born from pristine gas are hotter than metal-enriched (Pop~II) stars (and lack metal blanketing), and thus they have harder ionising continua (e.g., \citealt{Tumlinson_2000, Marigo_2001}). The large mass ($>100$\,\Msun) expected for Pop~III stars implies a very short lifespan (3--30~Myr). As they inevitably explode as supernovae, they are believed to pollute the surrounding pristine gas with specific chemical enrichment patterns, such as high C/O \citep[e.g.][]{vanni_characterizing_2023}, akin to the carbon-enhanced metal-poor stars in the Galactic halo. These high C/O chemical signatures have been potentially detected in a few objects \citep[e.g.][]{deugenio_jades_2023, scholtz_jades_2025-1, nakajima_ultra-faint_2025, Pollock2026}. 

Based on theoretical simulations, Pop~III stars are expected to form in small, low-mass haloes, or around more massive haloes at high redshift that are accreting pristine gas \citep[e.g.,][]{Abel_2002, Yoshida_2003, liu_integral_2020, Venditti_2023, Venditti_2024, Venditti_2025,Storck2026}. Direct observations of these objects are challenging due to their faint line and continuum emission \citep{Zackrisson_2015}. Therefore, a common approach is to exploit gravitational lensing, which boosts the intrinsically faint emission of these sources and increases the spatial resolution, significantly helping to find small stellar regions that are most likely to host Pop~III pockets. Combining this with the already excellent sensitivity and resolution of \jwst has enabled astronomers to find candidates for Pop~III galaxies \citep[e.g.,][]{trussler_2023, Adamo_2024, Fujimoto_2025}, with numerous oxygen-deficient candidates: AMORE6-B at $z = 5.725$ \citep[][]{Morishita_2025};  LAP1 at $z = 6.63$ \citep[][]{vanzella_extremely_2023, nakajima_ultra-faint_2025}; LAP-2 at $z=4.19$ \citep[][]{Vanzella_lap2_2026}. 
% and low-metallicity targets in \citet{Isobe_2026}. 
The absence or weakness of oxygen lines in these systems implies gas-phase metallicities below 1\%--5\% of the solar value, approaching the levels expected from self-pollution by Pop~III stars \citep{Katz_21, rusta_metal-polluted_2025}.
Even without gravitational lensing, we now have compelling PopIII candidates with HeII emission and lacking metal lines, in particular Hebe at $z=10.6$ \citep[][]{maiolino_jades_2024-1,  Maiolino_2026, Uebler_2026}.

At the same time, the extreme magnification provided by gravitational lensing for some sources has allowed the study of individual star-forming clumps or clusters in high-$z$ galaxies \citep[][]{Asada_2023, Claeyssens_2023, strait_extremely_2023, Adamo_2024, Bradac_2024, Mowla_2024, Bradac_2025,  Claeyssens_2025, Fujimoto_2025, Messa_2025, vanzella_pristine_2025,}. These clumps or clusters are found to have stellar masses in the range $10^{4}\text{--}10^{7}\,\Msun$, comprising more than 50\% of the total stellar mass of the host galaxy \citep[][]{Mowla_2024, vanzella_pristine_2025}, with stellar surface densities comparable to those of local star clusters \citep[][]{Fujimoto_2025}. Interestingly, many of the Pop~III candidates have masses similar to those of stellar clusters in the local Universe \citep[][]{Legus_2019} and at high-z. Given the similarities between the high-z stellar clusters and the Pop~III candidates, except for their low metallicities, it is likely that these Pop~III candidates are the progenitors of normal high-z stellar clusters or potentially the assembly of ultra-faint dwarfs \citep[][]{nakajima_ultra-faint_2025}.

In this work, we present new high-spectral-resolution ($R \sim 2700$) NIRSpec-IFS observations of LAP1 (Lensed and Pristine 1 at $z\sim6.6$; \citealt{Vanzella_2020}) covering the strong rest-frame optical emission lines (\OIIall to \Halpha; 2.7--5.3 $\mu$m observed frame), along with a complete re-analysis of the PRISM ($R \sim 100$) NIRSpec-IFS data \citep[covering \Lya to \Halpha;][]{vanzella_extremely_2023}, medium-resolution ($R \sim 1000$) NIRSpec-MSA data \citep[][]{nakajima_ultra-faint_2025}, and multi-band NIRCam imaging. LAP1 consists of two sources, dubbed A and B (\citealt{vanzella_extremely_2023}), with lensing producing double images for both A (A1, A2) and B (B1, B2). However, \jwst can spatially resolve only B1 and B2, so we treat A as the spatially integrated A1+A2. LAP1 lies on the caustic of the foreground lensing cluster, resulting in a magnification ($\mu_{\rm tot}$) of 120 for the B images and 500--20,000 for the A images (\citealt{Vanzella_2020, vanzella_extremely_2023}). The system is one of the most promising self-polluted Pop~III candidates, with a low \OIIIL/\Hbeta ratio ($\sim1.6$) and a non-detection of the UV or optical continuum, implying an upper limit on the stellar mass of $\lesssim10^{4} \Msun$. The previous detection of \CIVall by \citet{nakajima_ultra-faint_2025} indicates a hard ionising continuum, consistent with self-polluted Pop~III stars \citep{rusta_metal-polluted_2025}.
 
In \S~\ref{sec:obs} we present our observations and data reduction, and in \S~\ref{sec:analysis} we describe our emission line fitting. We present and discuss our results in \S~\ref{s.results}. Finally, in \S~\ref{s.conclusion} we summarise our results. Throughout this work, we adopt a flat $\Lambda$CDM cosmology: $H_0= 67.4$ km s$^{-1}$ Mpc$^{-1}$, $\Omega_\mathrm{m}$ = 0.315, and $\Omega_\Lambda$ = 0.685 \citep{2020A&A...641A...6P}. We use air wavelengths for the emission lines throughout the paper.

%--------------------------------------------------------------------
\section{Observations and data reduction}
\label{sec:obs}

The LAP1 arclet consists of the two objects LAP1-A and LAP1-B. Both A and B components are doubly imaged by the lensing cluster MACS J0416. While LAP1-B images (B1 and B2) are resolved, the images A1 and A2 of LAP1-A are unresolved on the critical line of the gravitational lens \citep[][]{vanzella_extremely_2023}. We show an overview of the system along with all available \jwst observations in Fig.~\ref{fig.overivew_obs}. Since component B is doubly imaged as B1 and B2, we combine the spectra of B1 and B2 into a single spectrum to increase the signal-to-noise ratio.

\subsection{NIRSpec-IFS Data}\label{s.IFS_obs}

We observed \target and its components with JWST/NIRSpec in IFS mode \citep{jakobsen_near-infrared_2022, boker_near-infrared_2022} as part of the GA-NIFS survey (PID 4528, PI: K. Isaak)\footnote{More information about the GA-NIFS survey can be found \href{https://ga-nifs.github.io/}{here}.}. The NIRSpec data were taken on 25\textsuperscript{th} January 2025, with a medium cycling pattern of twelve dither positions and a total integration time of 18.2~ks (5.05~h) with the high-resolution grating/filter pair G395H/F290LP, covering the wavelength range $2.87\text{--}5.27~\mu$m ($\lambda_{\rm rest}=3700\text{--}6900$\,\AA; spectral resolution $R\sim2000\text{--}3500$; \citealp{jakobsen_near-infrared_2022}). Additionally, in this work, we include PRISM/CLEAR IFS observations from GO PID 1908 (PI: Vanzella; taken on 16\textsuperscript{th}--17\textsuperscript{th} October 2022), with an on-source time of 22.3~ks, a wavelength coverage of $\lambda=0.6\text{--}5.3~\mu$m and a spectral resolution of $R\sim30\text{--}300$. A detailed description of the \prismifs observations is presented in \citet{vanzella_extremely_2023}.

Raw data files of these observations were downloaded from the Barbara A.~Mikulski Archive for Space Telescopes (MAST) and then processed with the {\it JWST} Science Calibration pipeline version 1.11.1 under the Calibration Reference Data System (CRDS) context \texttt{jwst\_1149.pmap}. We made several modifications to the default reduction steps to increase data quality, which are described in detail by \citet{perna_ga-nifs_2023} and which we briefly summarise here. Count-rate frames were corrected for $1/f$ noise through a polynomial fit. Furthermore, we removed regions affected by failed open MSA shutters during calibration in Stage 2. We also removed regions with strong cosmic ray residuals in several exposures. Any remaining outliers were flagged in individual exposures using an algorithm similar to {\sc lacosmic} \citep{van_dokkum_cosmic-ray_2001}: we calculated the derivative of the count-rate maps along the dispersion direction, normalised it by the local flux (or by three times the root mean square (rms) noise, whichever was higher), and rejected the 95\textsuperscript{th} percentile of the resulting distribution \citep[see][for details]{deugenio_fast-rotator_2024}. The final cubes were combined from the 2D files using the `drizzle' method. The main analysis in this paper is based on the combined cube with a pixel scale of $0.05''$.

Before we can analyse the emission line cubes of the PRISM or R2700 observations, we need to perform a number of data-preparation steps: 1) background subtraction; 2) masking of any outlier pixels that may have been missed by the pipeline; 3) flux uncertainty verification. For these tasks and the rest of the analysis, we use \texttt{QubeSpec}\footnote{Available on \href{https://github.com/honzascholtz/Qubespec}{GitHub}.}, an analysis pipeline written for NIRSpec/IFS data.

For both the R2700 and PRISM observations, we need to subtract the strong background affecting the data. For the R2700 observations, we mask the location of the source based on its \Halpha emission ($2\sigma$ SNR contours), and we estimate the background using the \texttt{astropy.photutils.background.Background2D} task with a $5\times5$ spaxel box window, for each individual channel in the data cubes to allow for spatial variation in the background. We visually inspected the resulting background spectra and found no evidence of narrow features (e.g. emission or absorption lines). Therefore, to reduce noise, we smoothed the background in spectral space using a median filter with a width of 25 channels. The final estimated background is subtracted from the flux data cube. We verified that the chosen background subtraction method does not influence the final conclusions.

We need to mask any major pixel outliers that were not flagged by the data reduction pipeline. Although these pixels do not cause significant problems during the emission line fitting of galaxy-integrated spectra, these outliers can become a problem during spaxel-by-spaxel fitting. To identify the residual outliers not flagged by the pipeline, we used the error extension of the data cube. We flagged any pixels whose error is more than $10\times$ the median error value of the cube. By testing different thresholds, we verified that this choice has no impact on the emission line maps or on our conclusions. 

A number of studies \citep[e.g.][]{ubler_ga-nifs_2023, scholtz_ga-nifs_2025} reported that the uncertainties on the flux measurements in the \texttt{ERR} extension of the data cubes are underestimated compared to the noise estimated from the rms of the spectrum, calculated inside a spectral window free from emission lines. However, the error extension still carries information about the relative uncertainties between pixels and outliers. Therefore, when extracting each spectrum, whether it is a combined spectrum of multiple spaxels or directly fitting a spaxel, we first retrieve the uncertainty from the error extension. Then we scale this error extension uncertainty so that the error extension's median uncertainty matches the spectrum's sigma-clipped rms in emission line-free regions. This scaling is performed independently for each detector, without wavelength dependence, for both the R2700 and PRISM IFS data. 

\subsection{NIRSpec-MSA Data}\label{s.MSA_obs}

Additional NIRSpec data of LAP1-B1 used in this work are from GO program ID 4750 (PI: K. Nakajima), taken using the microshutter array (MSA; \citealp{jakobsen_near-infrared_2022, ferruit_near-infrared_2022}). The target was observed in three visits on 4\textsuperscript{th} November 2025, using the medium-resolution gratings: G140M/F070LP and G395M/F290LP. This setup covers the nominal wavelength ranges of $0.9\text{--}1.3~\mu$m and $2.7\text{--}5.3~\mu$m with a resolution of ${\rm R\sim1000}$. The full description of the observations is presented in \citet{nakajima_ultra-faint_2025}. In this work, we reduced the data using the NIRSpec-GTO pipeline described in \citet{scholtz_jades_2025}. To obtain the total flux of the LAP1-B1 component, we used the 1D spectra extracted from an aperture of 5 pixels, corresponding to 0.5~arcsec, located at the target position in the 2D spectra.

\subsection{Astrometry alignment}\label{s.astrometry}

\begin{figure}
    \centering
    \includegraphics[width=0.99\columnwidth]{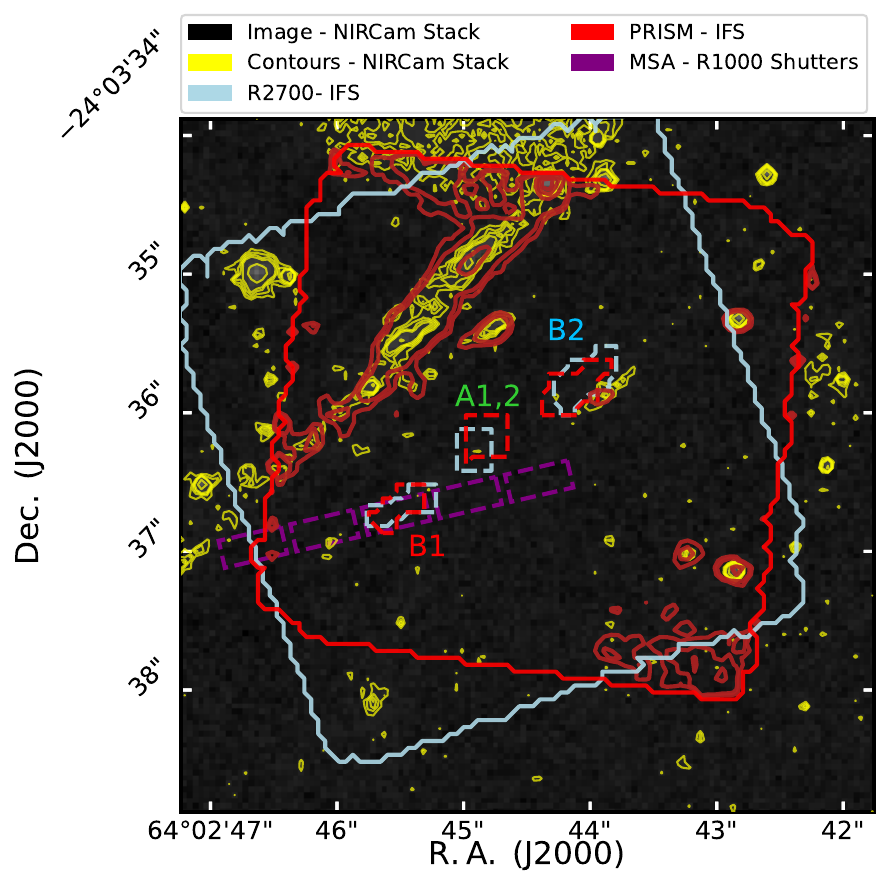}
    \caption{Overview of the observations targeting the LAP1 system and the astrometry alignment. We show the $5''\times 5''$ region around LAP1. Image: stack of all available NIRCam imaging from CANUCS \& PEARLS (F115W, F150W, F200W, F277W, F356W, F410M, F444W), with the yellow contours showing the $2, 3, 10, 15, 20\sigma$ levels. LAP1 is not detected in our stacked NIRCam imaging. The cyan and red dashed lines show the extraction regions for the R2700-IFS and PRISM-IFS observations used in this work, while the solid lines show the FoVs of the R2700-IFS and PRISM-IFS observations. The purple dashed rectangles show the location of the \MSA slit for the B1 component of LAP1.}
    \label{fig.overivew_obs}
\end{figure}

In this work, we combined four sets of \JWST observations of the LAP1 system: 1) stacked NIRCam imaging (see \S~\ref{s.masses}); 2) NIRSpec \MSA spectroscopy; 3) NIRSpec \prismifs; and 4) NIRSpec \highresifs observations. It is essential to align the astrometry of the different observations to confirm that we are extracting the spectra and photometry from the same regions of the LAP1 arc. While the NIRSpec-MSA observations were performed using the MSA target acquisition (MSATA) mode with multiple reference stars, the NIRSpec-IFS observations were performed without target acquisition, relying purely on the slew to the target, with a precision of $0.1\arcsec\text{--}0.3\arcsec$ \citep[][]{rigby_science_2023}. This is sufficient to fit a compact target into the IFU aperture, but additional alignment is required for a comparison with the NIRCam imaging. We assume that any offsets between the NIRCam imaging and \MSA observations are minimal, as our \MSA observations were performed with target acquisition. However, as the effective FoV of the NIRSpec/IFS observations is small ($2.7''\times2.7''$), there are no stars to correct the astrometry of the observations in the same way as for the NIRCam imaging. Instead, we use the crowded galaxy cluster field to manually align the astrometry between the PRISM-IFS and NIRCam imaging. We collapsed the PRISM-IFS cube to create a continuum map of the other objects within the FoV and matched the astrometry to the stacked NIRCam image. However, there is no continuum detected in the R2700-IFS observations, and therefore we used the detection of the \Halpha line in LAP1 (especially the B2 and A components; see \S~\ref{sec:eml_fit}) to align the astrometry between the NIRSpec/PRISM-IFS and R2700-IFS. We note that due to the limited SNR of the \Halpha line in both the \prismifs and \highresifs data, the accuracy of the \highresifs astrometry calibration is $\sim0.1$~arcsec. We show the final astrometric alignment of the observations on a NIRCam image in Fig.~\ref{fig.overivew_obs}.

\section{Data Analysis}\label{sec:analysis}

\subsection{Emission line fitting}\label{sec:eml_fit}

\begin{figure*}
    \centering
    \includegraphics[width=0.8\paperwidth]{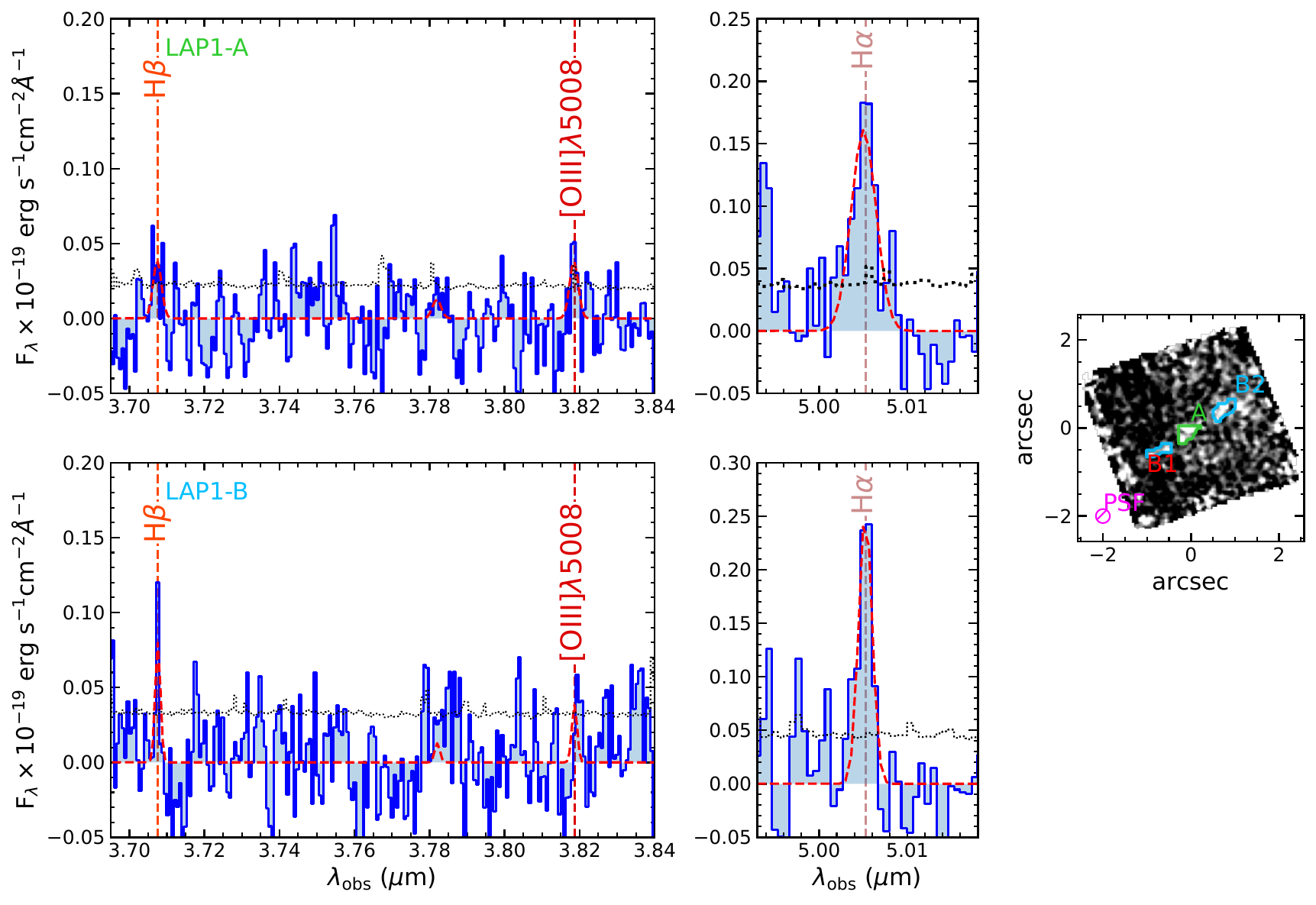}
    \caption{Overview of the NIRSpec \highresifs observations presented in this paper. Left two columns: \Hbeta, \OIIIall and \Halpha (left to right) continuum-subtracted spectra extracted from the A component and the sum of the B1+B2 components, with solid blue, dashed red and black dotted lines showing the data, best-fit model and 1$\sigma$ uncertainties, respectively. Right column: \Halpha emission line map, obtained by collapsing the channels around the \Halpha best fit, with contours showing the extraction regions of the spectra.}
    \label{fig.overivew_r2700}
\end{figure*}
\begin{figure*}
    \centering
    \includegraphics[width=0.8\paperwidth]{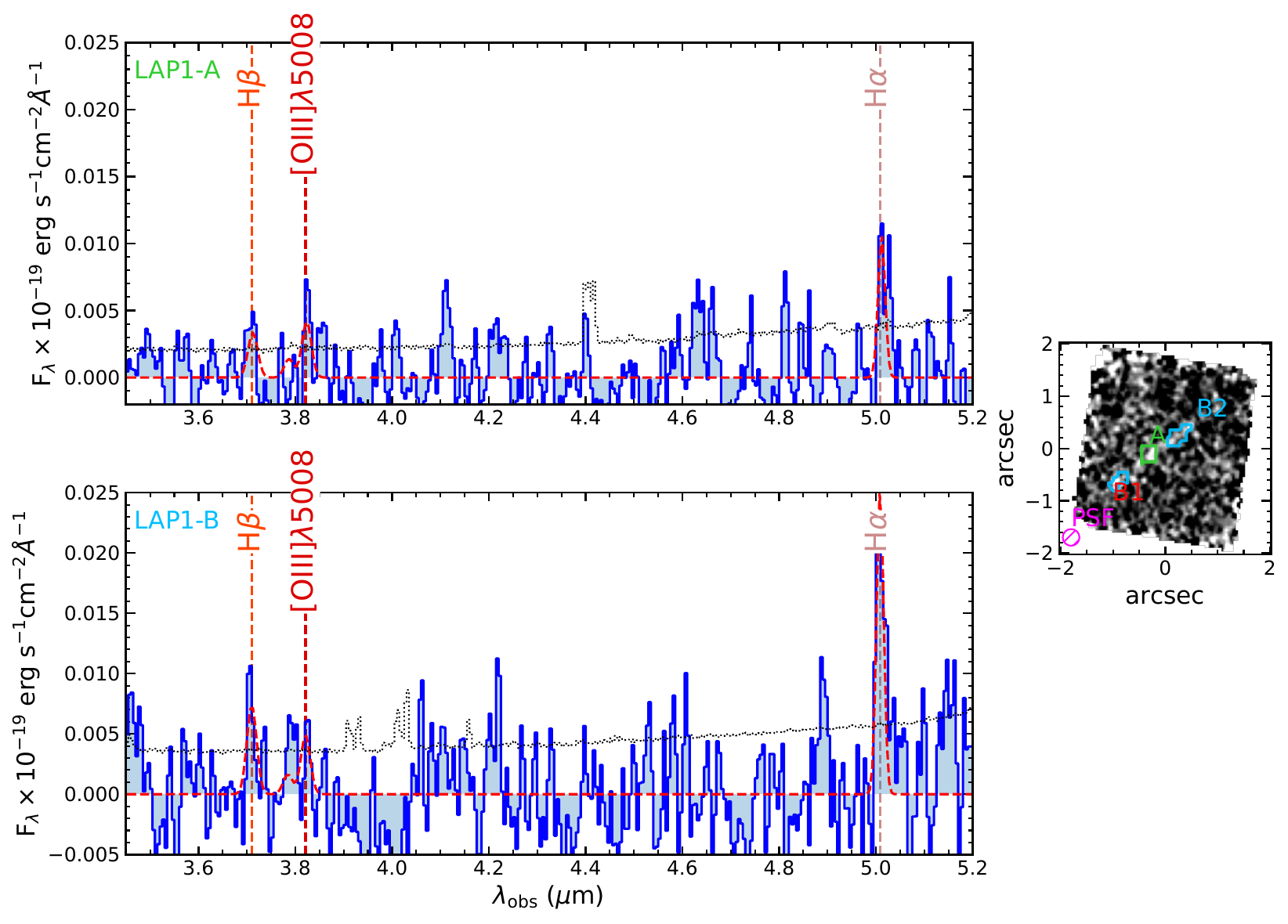}
    \caption{Overview of the NIRSpec \prismifs observations, as in Fig.~\ref{fig.overivew_r2700}. The right image is the \Halpha emission line map obtained by collapsing the channels around the \Halpha best fit. }
    \label{fig.overivew_prism}
\end{figure*}

We extracted the spectra based on the location of the three images originally identified in \citet{vanzella_extremely_2023} as LAP1-A, LAP1-B1, and LAP1-B2, as well as the total spectrum of LAP1-B by summing LAP1-B1 and LAP1-B2. We show the extracted spectra of LAP1-A and LAP1-B for \highresifs and \prismifs in Figs.~\ref{fig.overivew_r2700} and \ref{fig.overivew_prism}, respectively. We verified our results using the B1 and B2 spectra independently, but for the B component we only discuss the results from the summed spectrum. Each spectrum was fitted with a Gaussian profile for each emission line of interest: \OIIIall, \Halpha, \Hbeta and, in the case of the R1000 NIRSpec-MSA and PRISM-IFS data of LAP1-B1, also \CIVall, \Lya, \CIIIall, \HeIIL[1640] and \OIIIUvall. We note that, although we fitted \CIIIall, \HeIIL[1640] and \OIIIUvall, we do not detect any statistically significant ($>3\sigma$) emission. Additionally, we fitted the continuum as a power law; however, we note that the continuum was not detected in any of our observations, and this component primarily accounts for any residuals from the background subtraction. We use a single Gaussian per emission line as our main model, since we do not see any evidence for a broad component in any of the emission lines. We tie the redshift (centroid) and intrinsic FWHM of each Gaussian profile to a common value to reduce the number of free parameters, leaving the flux of each Gaussian profile free to vary. For each emission line, the FWHM of the line is convolved with the line spread function of NIRSpec from \citet{Shajib_2025}.
We fixed the \OIIIL[5008]/\OIIIL[4960] flux ratio to its theoretical ratio of 2.99 \citep{dimitrijevic_flux_2007}. For the PRISM-IFS data, we fixed the intrinsic FWHM of the emission line to 100 \kms\ (from the R2700 observations) as the resolution of the observations does not allow us to determine the line width. 

The fiducial model parameters of our best-fit models are estimated with a Bayesian approach, where the posterior probability distribution is calculated using the Markov chain Monte Carlo (MCMC) ensemble sampler -- \texttt{emcee} \citep{foreman-mackey_emcee_2013}. For each of the variables, we need to define priors for the MCMC sampling. The prior on the redshift of each spectrum is set as a truncated Gaussian distribution, centred on the systemic redshift of the galaxy with a sigma of 300~km~s$^{-1}$ and boundaries of $\pm 1000$~km~s$^{-1}$. The prior on the intrinsic FWHM is set as a uniform distribution between $10\text{--}500$~km~s$^{-1}$, while the prior on the amplitude of each line is set as a uniform distribution in log-space between 10\% of the rms of the spectrum and the maximum of the flux density in the spectrum. We verified that the posterior distributions do not reach any of the prior bounds outlined above. 

The final best-fit parameters and their uncertainties are calculated as the 50th percentile value and 68\% confidence interval of the posterior distribution. We note that all the quantities derived from our spectral fitting (e.g., gas-phase metallicities) are calculated from the posterior distribution to account for any correlated uncertainties in the spectrum. We summarise the measured fluxes from the \prismifs, \highresifs and \MSA observations in Table~\ref{tab.eml}. 

\begin{figure}
    \centering
    \includegraphics[width=0.99\columnwidth]{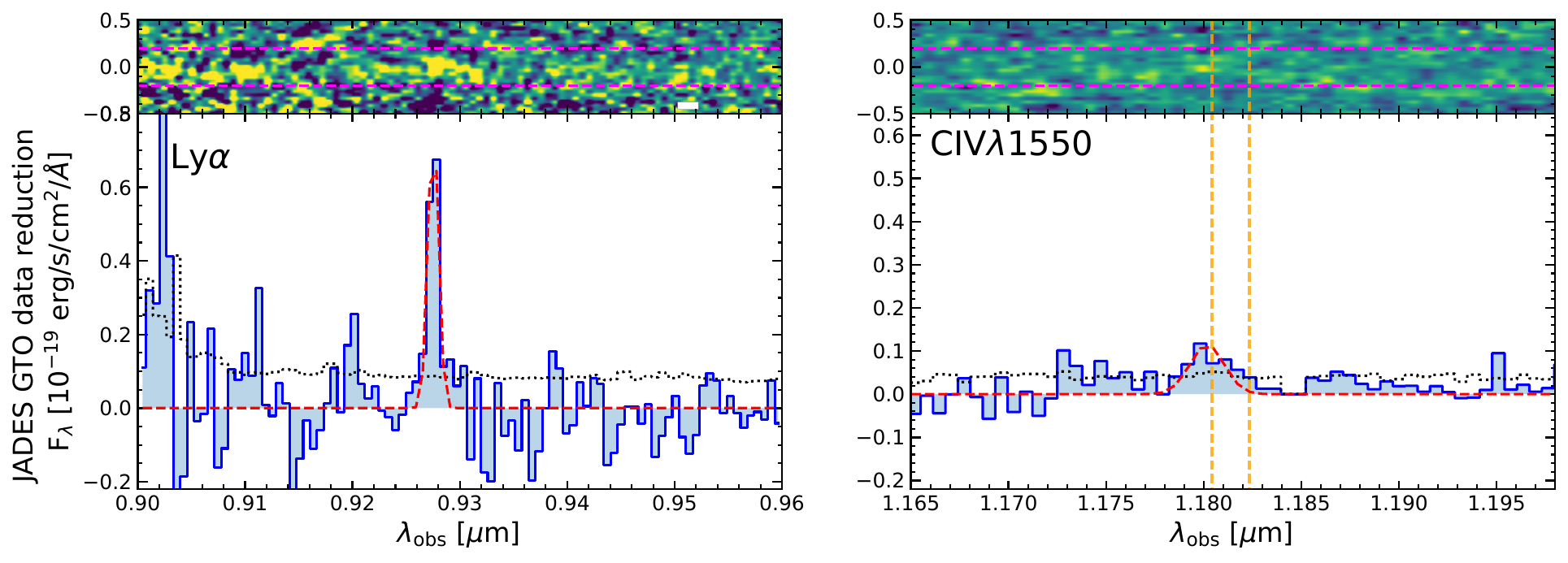}
    \includegraphics[width=0.99\columnwidth]{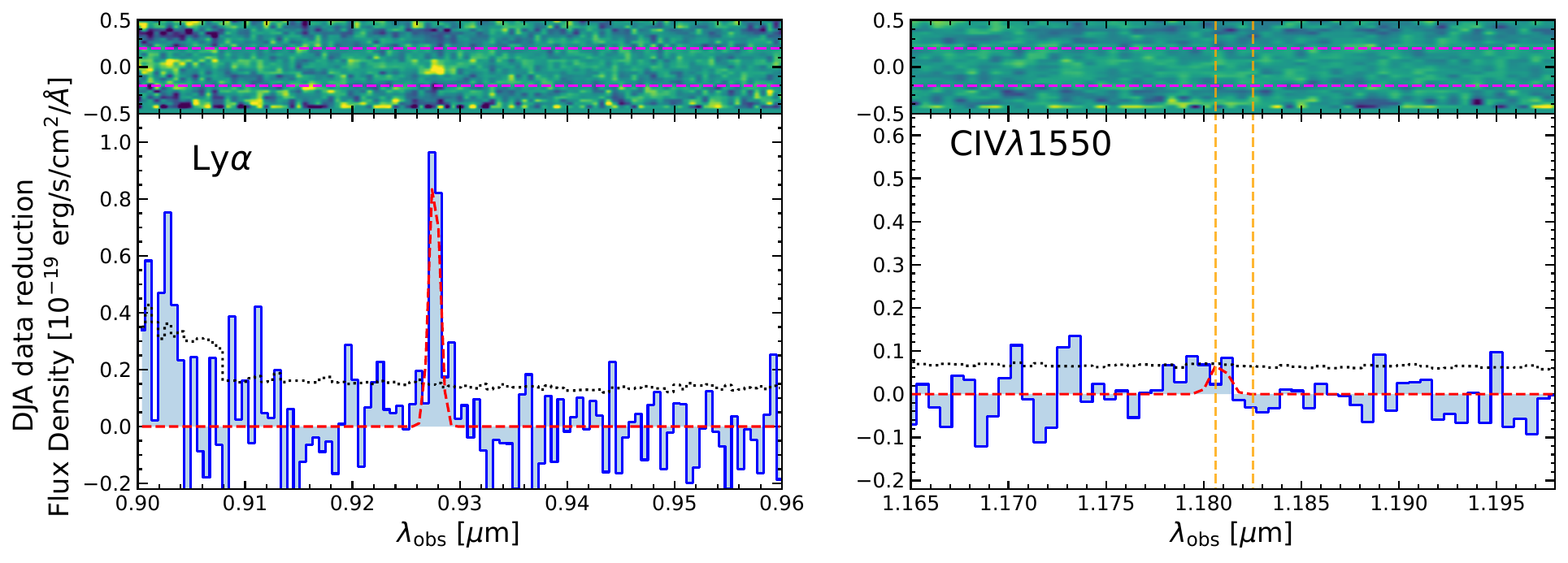}
    \caption{Spectra and best fits of the \Lya and \CIVall emission lines using our own re-reduction (and bootstrapped spectrum; top row) and DJA reduction (bottom row; see \S~\ref{s.MSA_obs}) of the \MSA data of the LAP1-B1 component. The solid blue, dashed red and black dotted lines show the data, best-fit model and the uncertainties, respectively. Above each spectrum, we show the 2D spectra from the MSA slit. We reliably detect \Lya in the \MSA, but we do not robustly detect \CIVall in either our reduction or DJA v4.}
    \label{fig.MSA}
\end{figure}

\section{Results \& Discussion}\label{s.results}

In this work, we combine the analysis of all available spectroscopy for the LAP1 system: \jwst/NIRSpec-IFS (PRISM and R2700 resolution) and \jwst/NIRSpec-MSA (R1000). Using the NIRSpec observations, we detect \Halpha in all components of LAP1 (A, B1 and B2), and we spectrally and spatially resolve the emission lines across all observations. Furthermore, we detect \Lya in all components of \target. We note that the \highresifs observations are not sufficiently deep to detect \Hbeta and \OIIIL[5008] at $>3\sigma$ in the individual A and B components. We therefore rely on the \prismifs and \MSA observations for metallicity estimates. However, we note that the \prismifs observations do not detect \Hbeta in the \target-A component, and hence we use the \Hbeta flux estimated from \Halpha, assuming Case B recombination and zero extinction, an assumption supported by the low metallicity. 

\subsection{Verifying \CIVall detection in LAP1-B}\label{s.c4_detection}

Previous works investigating the presence of Pop~III stars in the LAP1 system were based on the tentative detection of \HeIIL[1640] in \citet{vanzella_extremely_2023} (which was not confirmed in \citealt{nakajima_ultra-faint_2025} nor in this work) and \CIVall in \citet{nakajima_ultra-faint_2025}, which were considered evidence for the presence of a hard ionising continuum. The detection of \CIVall also suggested high C/O abundance, often associated with enrichment from Pop~III stars \citep{deugenio_jades_2023, scholtz_jades_2025-1}.

We show the \MSA spectra from the G140M and G395M gratings in Fig.~\ref{fig.MSA}, reduced using the NIRSpec-GTO pipeline (see \citealt{scholtz_jades_2025}). In order to provide additional verification of our data reduction, we also use the data from the Dawn \JWST archive (DJA), reduced using the \textsc{MSAexp} package \citep{brammer_msaexp_2023}, and we show the DJA spectrum in the bottom panel of Fig.~\ref{fig.MSA}. 

Using the JADES-GTO data reduction, we detect \Halpha, \Hbeta, \OIIIall and \Lya emission lines with measured fluxes within 1$\sigma$ of those measured by \citet{nakajima_ultra-faint_2025} (see Table~\ref{tab.eml}); however, we measure the \CIVall emission line with an SNR of 2.6$\sigma$ based on the MCMC fitting, compared to 3.1$\sigma$ in \citet{nakajima_ultra-faint_2025}. We further investigate the significance of the \CIVall emission by bootstrapping the individual exposures from the \MSA data. A full description of the method can be found in, e.g., \citet{maseda_jwstnirspec_2023, deugenio_jades_2024, hainline_searching_2024, curti_jades_2024, scholtz_jades_2025-1}. We generated 5000 bootstrapped spectra by randomly sampling (with replacement) over the set of individual sub-spectra (i.e. spectra extracted from each of the 40 individual exposures), after four passes of iterative 3$\sigma$-clipping to remove any remaining outliers not flagged by the pipeline. For each random realisation of the spectrum, we fitted the model described above to estimate the continuum shape and measured fluxes. The final measured fluxes of the emission lines from the bootstrapping approach are the means of the distributions, with the error estimated as the standard deviation. These bootstrapped uncertainties are considered to be more accurate than the flux density uncertainty estimates from the pipeline as they consider all sources of noise, including correlated noise from resampling the NIRSpec spectra. We find a consistent non-detection of the \CIVall emission line (2.1$\sigma$, although this can be considered at tentative detection), while the fluxes and SNRs of the other emission lines are consistent across data reductions and with \citet{nakajima_ultra-faint_2025}, who reduced the data using the standard STScI pipeline. As a result, while our inferred \CIVall flux is statistically consistent with the value reported by \citet{nakajima_ultra-faint_2025}, we cannot confirm the detection at the $3\sigma$ level. This is most likely due to minor differences in the pipelines, along with our bootstrapping of the individual exposures to robustly sample all sources of uncertainties in the NIRSpec observations. 

\begin{figure}
    \centering
    \includegraphics[width=0.99\columnwidth]{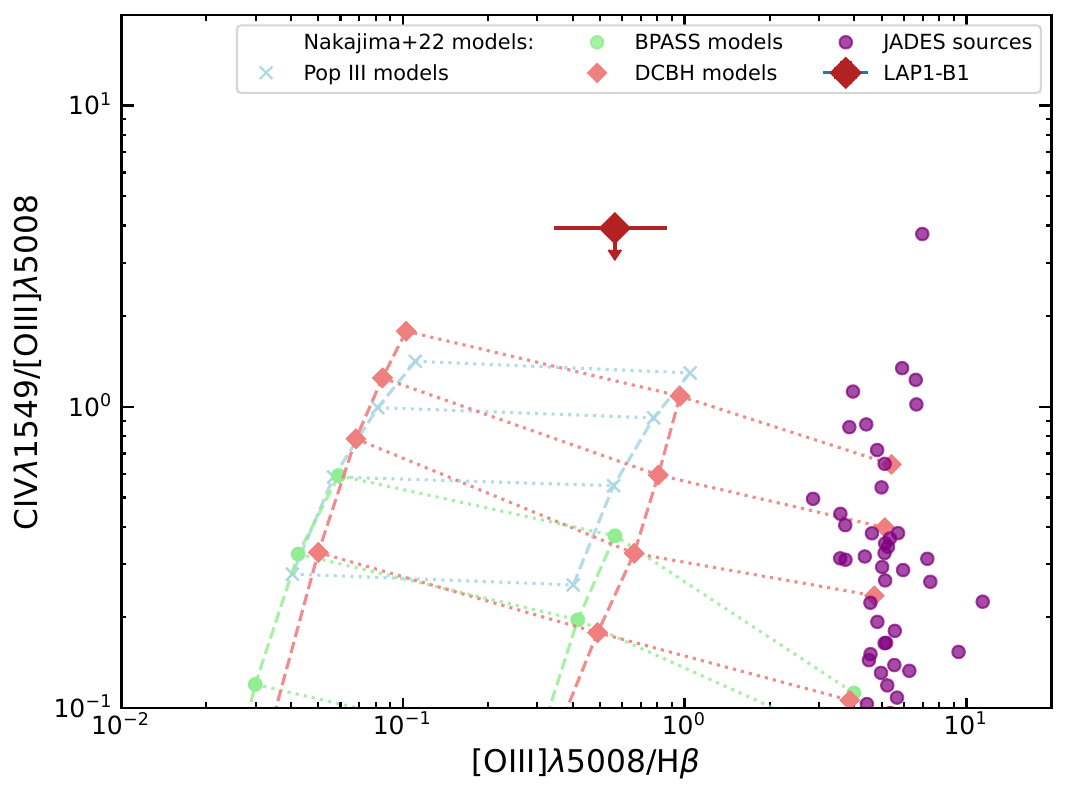}
    \caption{\CIVall/\OIIIL[5008] vs \OIIIL[5008]/\Hbeta tracing the metallicity and ionisation parameter. We show the LAP1 measurement as a red diamond and compare our measurement to the models from \citet{nakajima_diagnostics_2022} for Pop~III models (blue crosses), stellar models using BPASS (green circles) and direct collapse black holes (DCBHs; salmon diamonds). The dark magenta circles show \CIVall detections in JADES DR4 \citep[][]{scholtz_jades_2025}.}
    \label{fig.C4O3_plot}
\end{figure}

In Fig.~\ref{fig.C4O3_plot}, we plot \CIVall/\OIIIL[5008] vs \OIIIL[5008]/\Hbeta, where the y-axis traces the ionisation and the C/O abundance and the x-axis traces the metallicity. The non-detection of \CIVall and the faint detection of \OIIIL[5008] result in an upper limit on \CIVall/\OIIIL[5008], which is consistent with all possible low-metallicity models \citep[][]{nakajima_diagnostics_2022}, including Pop~III models, typical evolved star-forming models, and direct-collapse black hole (DCBH)/AGN models. Therefore, with this diagram, we cannot firmly confirm the presence of Pop~III stars in the LAP1-B object. However, the 2\,$\sigma$ signal is very encouraging, and future observations are necessary to firmly establish or reject this consequential detection.

\subsection{Oxygen and Carbon abundances}\label{s.abud}

\begin{figure}
    \centering
    \includegraphics[width=0.99\columnwidth]{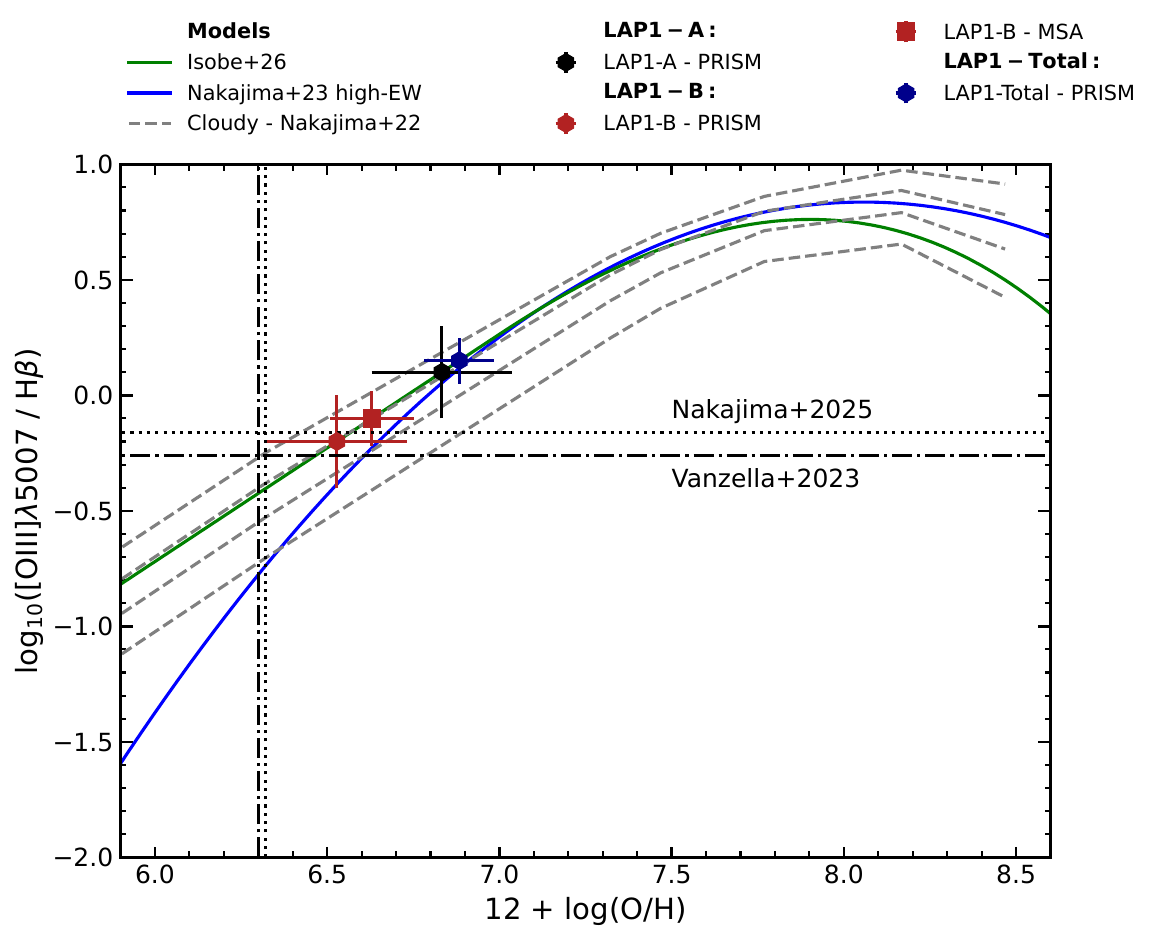}
    \caption{Estimate of the metallicity (oxygen abundance) based on the \OIIIL[5008]/\Hbeta (R3) ratio. We show the measurements for LAP1-A, LAP1-B and total as black, red and blue points, respectively.
    We show various calibrations between R3 and metallicity: Cloudy models (\citealt{nakajima_diagnostics_2022}, dashed lines), high emission-line EW (\citealt{nakajima_jwst_2023}, blue solid line) and \citet{Isobe_2026} calibration based on low-metallicity sources (green line; our fiducial choice). The dotted and dash-dotted lines show the previous measurements for LAP1-B1 from \citet{nakajima_ultra-faint_2025} and \citet{vanzella_extremely_2023}, respectively. Our metallicity estimates are consistent with previous measurements when the same calibration is used.}
    \label{fig.metal_calib}
\end{figure}

In order to measure the metallicity of \target and its components, we use the ratio $R3 = \log(\OIIIL[5008]/\text{H}\beta)$, using various calibrations from the literature: 1) \texttt{Cloudy} photoionisation modelling from \citet{nakajima_diagnostics_2022}; 2) the calibrations from \citet{nakajima_jwst_2023}; and 3) new calibrations for low oxygen abundance targets from \citet{Isobe_2026}. We show these calibrations together in Fig.~\ref{fig.metal_calib}. We adopt the calibrations from \citet{Isobe_2026} for our fiducial measurement, as these are the direct, empirical calibrations at high-z that extend to the lowest metallicity, hence requiring the least extrapolation. We measure an oxygen abundance, $12+\log(\text{O/H})$, of 6.9, 6.8 and 6.5 for LAP1-total, LAP1-A and LAP1-B, respectively. We note that our oxygen abundances are, on average, 0.2 dex higher than the previously measured values from \citet{vanzella_extremely_2023} and \citet{nakajima_ultra-faint_2025} despite our R3 measurements being consistent with those from the previous studies. This is due to our choice of using abundance calibrations derived from \jwst observations of high-z, low-metallicity galaxies \citep[][]{Isobe_2026} rather than assuming calibrations derived from high-ionisation-parameter models as done in previous studies. 
% We choose this calibration due to the non-detection of either \HeIIL[1640] or \CIVall (see \S~\ref{s.c4_detection} above). 

\begin{figure*}
    \centering
    \includegraphics[width=0.8\paperwidth]{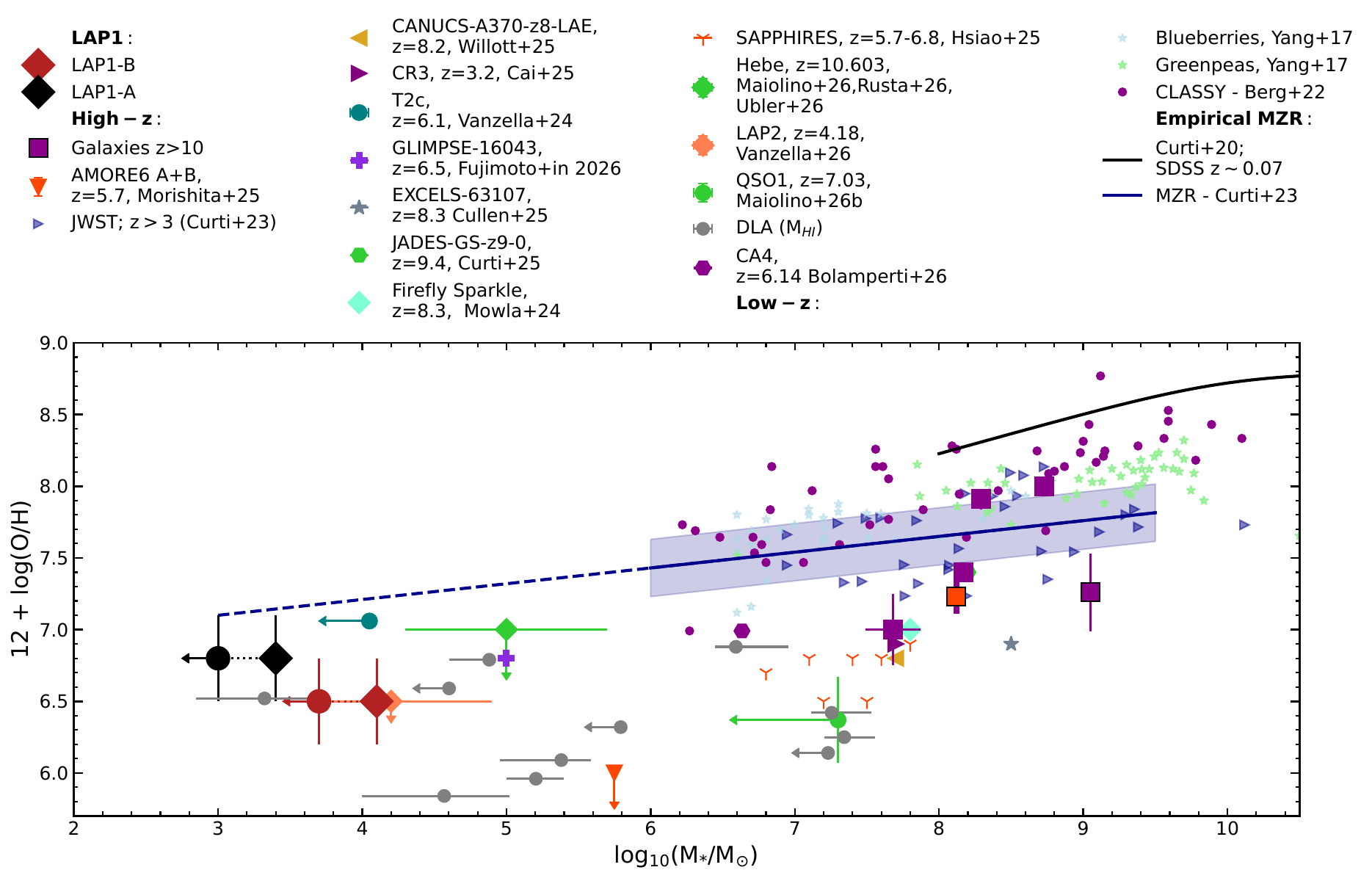}
    \caption{Oxygen abundance vs stellar mass for LAP1-A and LAP1-B. The diamond and circle points indicate stellar mass estimates from the \Hbeta and UV continuum, respectively. Additional measurements of low-metallicity galaxies and AGN at $z>3$ from the literature are reported \citep{vanzella_extremely_2023, Morishita2025_lowZ, Mowla_2024, Tripodi_nat_2025, vanzella_extreme_2024, Hsiao2025_lowZ, curti_jades_2025, cullen_jwst_2025, Willott_lowz_2025, maiolino_jades_2024-1, nakajima_ultra-faint_2025, Cai_2025, Morishita_2025, Uebler_2026, Rusta_2026,Bolamperti_2026} and the Blueberries and Green Peas (local analogues of high-z galaxies; \citealt{yang_blueberry_2017}). We show the MZR for z$>$3 from \citet{curti_jades_2023} as a blue solid line (the dashed line indicates extrapolation to low stellar mass).}
    \label{fig.metal_mass}
\end{figure*}

We locate the LAP1 components on the mass--metallicity relation (MZR)  in Fig.~\ref{fig.metal_mass} and compare them to other low-metallicity sources at high-z from the literature, using the stellar masses estimated in \S~\ref{s.masses}. Our results confirm that \target-B is the most metal-poor component. However, the absolute value of the oxygen abundance is subject to a $\sim$0.3~dex systematic error due to uncertainties in the metallicity calibrations. 

\begin{table}
   \caption{Emission line fluxes of the three main components identified in the literature from the PRISM, R2700 and R1000 data. The fluxes are in units of $10^{-19}$ \ergscm. For each flux and error, we include the estimated SNR in parentheses.}
   \centering
 \begin{tabular}{lccc} 
\hline 
\hline 
  &   R2700-IFS & PRISM-IFS & \MSA \\
\hline 
LAP1-Total\\
\hline
\Halpha& 11.1$\pm$1.3 (8.5)& 9.1$\pm$0.9 (10.1) & - \\
\OIIIL[5008]& 3.2$\pm$0.6 (5.3)& 3.2$\pm$0.6 (5.3)& -\\
\Hbeta & 2.9$\pm$0.6 (4.8) &  3.3$\pm$0.7 (4.7)& -\\
\Lya & -  & 22.0$\pm$4.5 (4.9)&  -\\
R3  & - & 0.15$\pm$0.2 & - \\
12+log(O/H) &-  & 6.9$\pm$0.1 & - \\
FWHM [\kms] & 181$^{+30}_{-30}$ & - & - \\
\hline 
LAP1-A\\
\hline
\Halpha& 5.6$\pm$0.8 (7.0)&  1.93$\pm$ 0.47 (4.1) & - \\
\OIIIL[5008]& 1.09$\pm$0.37 (2.9)& 1.19$\pm$0.24 (4.95) & -\\
\Hbeta & 1.07$\pm$0.38 (2.8) &  0.70$\pm$0.28 (2.5)& -\\
\Lya & -  & 14.0$\pm$2.3 (6.1) &  -\\
R3  & - & 0.1$\pm$0.2 & - \\
12+log(O/H) & - & 6.8$\pm$0.2 & - \\
FWHM [\kms] &  180$^{+37}_{-35}$  & - & - \\
\hline
LAP1-B1\\
\hline
\Halpha&2.3$\pm$0.4 (5.8)& 1.9$\pm$0.2 (9.5)& 2.0$\pm$0.2 (10.0)\\
\OIIIL[5008]&0.3$\pm$0.2 (1.5)& 0.7$\pm$0.2 (3.5)& 0.6$\pm$0.1 (6.0)\\
\Hbeta & 0.75$\pm$0.25 (3.0)& 0.9$\pm$0.2 (4.5)& 0.7$\pm$0.2 (3.5)\\
\CIVall & - & - & 1.7$\pm$0.7 (2.1)  \\
\Lya & - & 6.1$\pm$1.9 (3.2) &  9.1$\pm$1.0 (9.1) \\
R3 & -& -0.1$\pm$0.2 & -0.2$\pm0.1$  \\
12+log(O/H) & - & 6.5$\pm$0.2 &6.6$\pm$0.2  \\
FWHM [\kms] & 71$^{+28}_{-36}$  & - & 100$^{+54}_{-49}$ \\
\hline
LAP1-B1,2\\
\hline
\Halpha& 4.9$\pm$0.6 (8.1)& 4.7$\pm$ 0.7 (6.7)& -\\
\OIIIL[5008]& 0.72$\pm$0.35 (1.9)& 1.3$\pm$0.4 (3.1)& -\\
\Hbeta & 1.4$\pm$0.4 (3.5)& 1.9$\pm$0.5 (3.8)& -\\
\Lya & - & 12.4$\pm$2.7 (4.59) &  - \\
R3 & - & -0.2$\pm$0.2 &  -\\
12+log(O/H) & - & 6.5$\pm$0.2 & - \\
FWHM [\kms] & 64$^{+22}_{-26}$  & - & - \\
\hline
\hline
\end{tabular} 
  \label{tab.eml}
\end{table}

\begin{table}
   \caption{Derived properties of the A and B components in the LAP1 system. The upper limits are based on 3$\sigma$ limits.}
   \centering
 \begin{tabular}{lcc} 
\hline 
\hline 
 Property &   LAP1-A & LAP1-B \\
\hline 
\Halpha radius [pc] & <10  & <10  \\
$\log_{10}(M_{\rm dyn}/\Msun)$(10pc) & $<7.6^{+0.2}_{-0.2}$ & $<6.6^{+0.3}_{-1.2}$ \\
$\log_{10}(M_{\rm dyn}/\Msun)$(1pc) & $<6.6^{+0.2}_{-0.2}$ & $<5.6^{+0.3}_{-1.4}$ \\
$\log_{10}(M_{*}/\Msun)$ & $<$3.0 & $<$3.7 \\
$12+\log(\text{O/H})$ & 6.8$\pm$0.2 &  6.5$\pm$0.2  \\
$\log_{10}(\text{C/O})$ & - & $<$-0.06 \\
$\log_{10}(\ksio/\text{[Hz/erg]})$ &  >26.4& >26.0 \\
\fescLya & 0.8$\pm$0.3 &  0.5$\pm$0.1 \\
$\log_{10}(\text{SFR}_{\Halpha}/[\msun\,\text{yr}^{-1}])$ & -2.4 & -2.2 \\
\hline
\hline
\end{tabular} 
  \label{tab.prop}
\end{table}

\begin{figure}
    \centering
    \includegraphics[width=0.99\columnwidth]{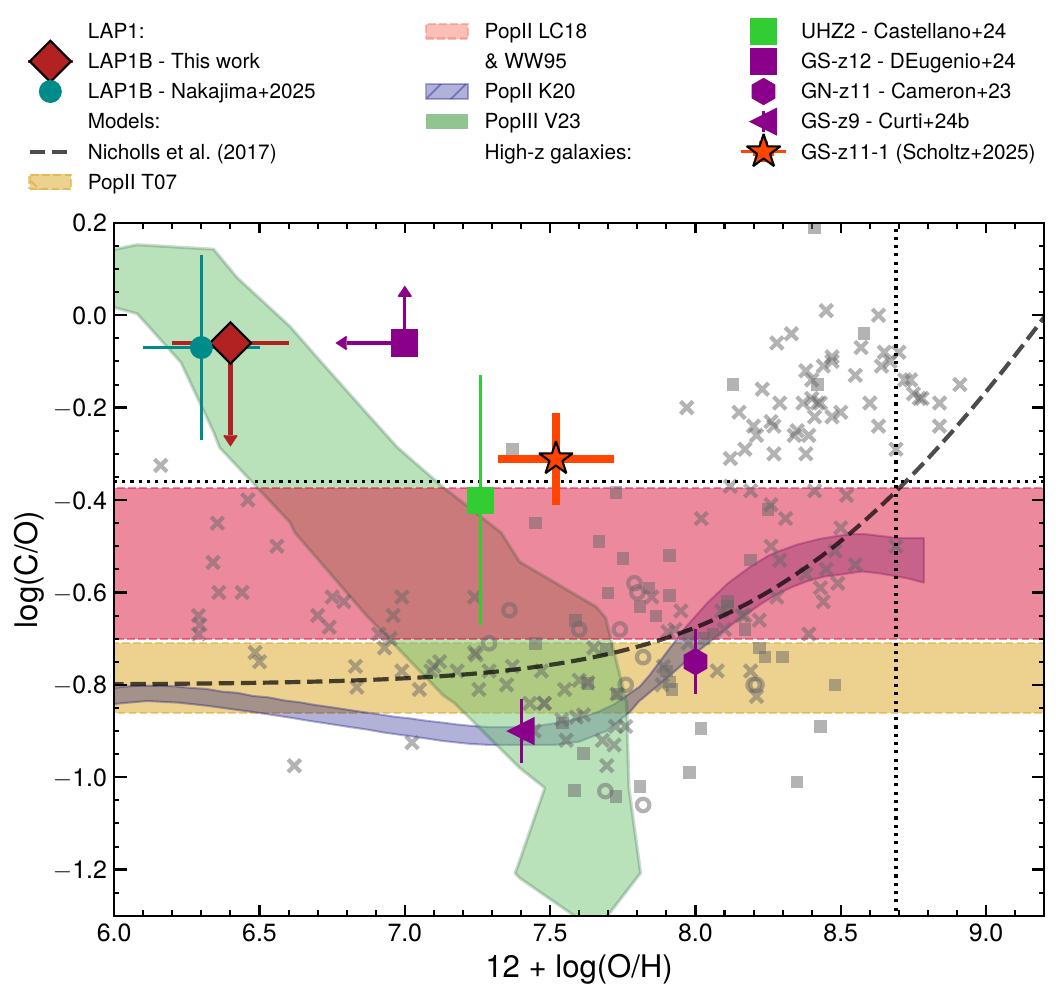}
    \caption{C/O vs O/H for LAP1-B (red star; the measurement from \citealt{nakajima_ultra-faint_2025} is shown as a cyan point), compared with literature values for $z>9$ galaxies \citep[][]{cameron_jades_2023-1, curti_jades_2024-1, deugenio_jades_2023, castellano_jwst_2024, scholtz_jades_2025-1}. We show theoretical yields of Pop~II \& III supernovae (black dashed line; yellow, red, blue and green shaded regions; \citealt{woosley_evolution_1995, tominaga_supernova_2007, heger_nucleosynthesis_2010, limongi_presupernova_2018, vanni_characterizing_2023}). The black dotted line shows the solar abundances as defined by \citet{gutkin_modelling_2016}.}
    \label{fig.CO_abundance_plot}
\end{figure}

We estimate the C/O abundance of \target-B1, following the method described in \citet{nakajima_ultra-faint_2025} using the \CIVall and \OIIIL[5008] flux measurements and upper limits. Given the non-detection of \CIVall, we are only able to estimate an upper limit on C/O. We estimate $\log_{10}(\text{C/O})<-0.06$, and we compare our estimated value to other targets in the literature and models in Fig.~\ref{fig.CO_abundance_plot}. Given the upper limit on C/O, we find that the low metallicity can be reproduced by both Pop~III and Pop~II stars. Yet, if the tentative $2\sigma$ signal in CIV were confirmed, then this would locate Lap1-B in the PopIII regime. We discuss futher the nature of LAP1-B and LAP1 in \S~\ref{s.discussion}.

\subsection{Stellar \& Dynamical Masses}\label{s.kinematics}

\subsubsection{Stellar masses}\label{s.masses}

The LAP1 system is not detected in any of the NIRCam filters, despite combining deep NIRCam observations from the PEARLS and CANUCS imaging surveys (PID:1218, 1176 \& 2738). Therefore, we mean-stacked all the filters covering the rest-frame FUV--optical range of LAP1: F115W, F150W, F200W, F277W, F356W, F410M and F444W. We created three stacked images: (i) stacking all filters together; (ii) stacking the filters covering the rest-frame UV (F115W, F150W, F277W); and (iii) stacking the filters covering rest-frame optical wavelengths (F356W, F410M, F444W). We do not detect LAP1 in any of our stacked images. The final stacked image comprises $\sim$45 hours of integration time in total. 

As in previous analyses of the LAP1 system \citep[][]{vanzella_extreme_2024, nakajima_ultra-faint_2025}, we do not detect any component of the arclet in the stacked data. We estimate a 3$\sigma$ upper limit on the flux density by bootstrapping mock apertures on the stacked image after masking detected sources, as this is a more conservative approach to estimate the noise in the data. We masked all detected emission at $>3\sigma$ using the \texttt{photutils.background} routine. We then placed 1000 random apertures with the size of LAP1-A and LAP1-B as seen in line emission ($r=0.2$~arcsec) and estimated the noise as the rms of the resulting distribution. 

We place a 3$\sigma$ upper limit on the flux density of 1.8~nJy for an aperture with a radius of 0.2~arcsec, corresponding to an apparent magnitude of 30.5 (delensed apparent magnitudes of $>35.7$ and $>37.3$~mag for the B and A components, respectively). The non-detection of the optical continuum in this system indicates a high equivalent width of the emission lines -- an \Halpha EW of $>1200\,\text{\AA}$, indicative of a very young stellar population of $<5$~Myr. In order to estimate the stellar mass of the components, we follow the method 
outlined in \citet{vanzella_extreme_2024} using the Starburst99 library \citep[][]{Leitherer_1999}, assuming a Salpeter IMF \citep[][]{shetty_salpeter_2014} and an instantaneous burst scenario at the lowest available metallicity (5\% Z$_{\odot}$) with an age younger than 5~Myr (based on the EW of the \Halpha line). We estimate the stellar mass both from the non-detection of the UV continuum and from the stellar mass required to produce the \Halpha emission. However, we note that for such a low-mass system, the stellar mass estimates may be significantly affected by stochastic sampling of the IMF \citep[see e.g.][]{Rossi_21}.

Based on the UV continuum, we measure stellar masses of $\log(M_{\star}/\Msun)<3.0$ and $<3.7$ for LAP1-A and LAP1-B, respectively. The upper limits on the stellar mass are consistent with those found in \citet{vanzella_extremely_2023} and \citet{nakajima_ultra-faint_2025} ($\log(M_{\star}/\Msun)<3-4$). Based on the line emission, we estimate that such a young stellar population produces an \Hbeta luminosity of $10^{34.4}$~erg/s/\Msun. This corresponds to a stellar mass of $10^{4.1}\,\Msun$ for LAP1-B and $10^{3.4}\,\Msun$ for LAP1-A. However, we note that for LAP1-A, the largest uncertainty is the lensing magnification of the source, which we discuss further in \S~\ref{s.discussion}.

There is a 0.4 dex discrepancy between H$\beta$ and UV continuum stellar mass estimates. This could indicate that the ages are even younger than the assumed 2--5~Myr or that high-mass stars are present, boosting the emission lines relative to the UV continuum. However, detailed modelling of the UV continuum is beyond the reach of the data currently available and should be explored with future observations and facilities.

\subsubsection{Dynamical masses}

We spectrally resolved the \Halpha emission line in the \highresifs data, measuring the FWHM of the line for the A and B components as $180^{+37}_{-35}$ and $64^{+22}_{-26}$~\kms, respectively. The FWHM in the \highresifs data is a factor of two lower than in the \MSA data, consistent with reports of R1000 observations overestimating the line width compared to the R2700 observations \citep{juodzbalis_jades_2025}. 

We do not detect any significant velocity gradient in the data that would indicate rotation or any other ordered motion of the gas. We investigated the presence of a velocity gradient in either of the components with two separate methods: 1) creating \Halpha emission line maps of the red and blue sides of the emission line; 2) extracting spectra from two regions along each component. Both methods show a hint of a velocity gradient of $\sim10$~\kms\ in all three sources (A, B1 and B2; see Fig. \ref{fig.vel_grad} and \S~\ref{s.app.kin}). However, we attribute this velocity gradient to known inaccuracies in the NIRSpec-IFS instrument model%
\footnote{More information at \url{https://www.stsci.edu/files/live/sites/www/files/home/jwst/documentation/technical-documents/_documents/JWST-STScI-009029.pdf}}
for the following reasons: a) the velocity gradients of B1 and B2 are in the same direction, even though gravitational lensing should mirror them across the critical line; and b) all three velocity gradients are aligned perpendicular to the slicer direction. We confirm that the spurious velocity gradient does not have a significant effect on the measured line width. We note that the expected velocity gradient in a 1~pc slit in simulations of star clusters (see discussion below) is $<20$~\kms\ \citep[][]{Lahen_2025}. Therefore, the expected velocity gradient is of a similar order to the current instrumental miscalibration; however, measuring it could in principle become achievable with an improved instrument model in future iterations of the \jwst pipeline and with deeper observations.

Given the lack of a physical velocity gradient in the data, we use the velocity dispersion and the \Halpha sizes of the individual clumps to measure the dynamical masses, assuming that the gas in these sources is virialised. We estimated the dynamical mass following the approach described by \citet{ubler_ga-nifs_2023}:
\begin{equation}
    M_{\rm dyn} = \frac{\kappa\sigma^{2}R_{\rm e}}{G},
\end{equation}
where $\sigma$ is the integrated velocity dispersion, $\kappa$ is the coefficient accounting for the mass distribution in the object (here assumed to be 5 as it is the most conservative value; \citealt{cappellari_sauron_2006}), and $R_{\rm e}$ is the effective radius of the sources. When using the velocity dispersion of the gas in such a low-mass system, it is important to take into account the thermal broadening of the line. For an ISM temperature of $2\times 10^{4}$~K, we would expect a thermal broadening of $\sigma_{\rm thermal}\sim 14$~\kms. This thermal broadening must be subtracted from the measured velocity width in quadrature to estimate the velocity dispersion due to gas motion.

One of the largest uncertainties for the dynamical masses is the physical size of the A and B components. We note that in both cases the \Halpha emission appears to be larger than the PSF of NIRSpec in both IFS and MSA modes. However, the low SNR of the observations does not allow us to reliably measure the size of either component. The extended nature of the \Halpha emission is further confirmed in the MSA data, where the \Halpha flux in the 5-pixel extraction is a factor of 2 lower than in the 3-pixel extraction, implying that the \Halpha is spatially resolved in the slit. Combining the IFS and MSA data, we conclude that the \Halpha emission is extended on a scale of 0.3--0.5~arcsec. However, the dominant source of uncertainty is the systematic uncertainty in the lensing model \citep[see][]{Meneghetti_2017, vanzella_extremely_2023}. We adopt clump sizes of $<$10~pc for the A and B components, based on an estimated upper limit on the size of a few pc to a few tens of pc \citep{vanzella_extremely_2023}. We measure dynamical masses of $\log_{10}(M_{\rm dyn}/\Msun)$  $<7.6^{+0.2}_{-0.2}$ and $<6.6^{+0.3}_{-1.2}$ for the A and B components, respectively. The uncertainties on the upper limits reflect the 1$\sigma$ uncertainties on the measured FWHM of the emission line. However, we note that these are conservative upper limits due to the large assumed size of the objects. Assuming that this is an early stage of a star cluster (see discussion in \S~\ref{s.discussion}), the expected sizes of these systems would be $<1$~pc (e.g. \citealt{Brown_gnedin_2021}, \citealt{Adamo_2024}, \citealt{Vanzella_lap2_2026}, and theoretical work by \citealt{Lahen_2025}), reducing the dynamical mass estimate by a factor of 10. This would imply $\log_{10}(M_{\rm dyn}/\Msun) <6.6^{+0.2}_{-0.2}$ and $<5.6^{+0.3}_{-1.4}$ for the A and B components, respectively. Another factor affecting $M_{\rm dyn}$ is the assumption that the ionised gas is virialised, which also makes the inferred value an upper limit, although this effect is challenging to quantify.

\subsection{\Lya properties}\label{s.Lya}

We leverage the detection of \Lya and \Halpha in the individual components to estimate the \Lya escape fraction (\fescLya). The ratios of \Lya/\Halpha for the A and B components are 7.3$\pm2.9$ and $4.5\pm0.7$, respectively. In order to estimate the escape fraction of \Lya photons ($f_{\rm esc}^{\Lya}$), we adopt the canonical ratio $\Lya/\Halpha = 8.7$ \citep[e.g.,][]{Henry_2015} to calculate:
\begin{equation}
    f_{\rm esc}^{\Lya} = \frac{\Lya}{\Halpha \times 8.7 }
\end{equation}
assuming Case B recombination with $T_{\rm e} = 2\times 10^4$~K and $n_{\rm e} = 100$~cm$^{-3}$ \citep[][]{osterbrock_astrophysics_2006}. We obtain \fescLya\  of $0.8\pm0.3$ and $0.5\pm0.1$ for the A and B components, respectively. We note that we do not correct for dust attenuation as we do not see any evidence for dust since the \Halpha/\Hbeta ratio is consistent with Case B (see Table~\ref{tab.eml}). Large \fescLya\ values ($>30$\%) have been observed in high-z \Lya emitters (LAEs; e.g. \citealt{Hayes_2011, Steidel_2011, saxena_jades_2023}), often in galaxies with very blue UV continuum slopes \citep[e.g.][]{Tang_2024, Steidel_2011, Bolamperti_2026}. Similarly high \fescLya\ values, reaching $\sim$30--50\%, have been observed in confirmed LyC-leaking galaxies at $z<3$ (e.g. \citealt{izotov_j08114730_2018, Smith_2020, Marques-chaves_2022}).

With the upper limit on the UV continuum and \Halpha line, we estimate the hydrogen-ionising photon production efficiency, \ksi, as
\begin{equation}
   \xi_{\rm ion} = \frac{\dot N_{\rm ion}}{L_{\nu}^{\rm UV}},
\end{equation} 
where $\dot N_{\rm ion}$ (s$^{-1}$) is the intrinsic production rate of hydrogen-ionising photons from stellar populations and $L_{\nu}^{\rm UV}$ (erg~s$^{-1}$~Hz$^{-1}$) is the (monochromatic) UV continuum luminosity, for which we derived upper limits from the stacked blue NIRCam filters. \ksi\ intrinsically depends on the assumed stellar-population model \citep[e.g.,][]{robertson13, eldridge17, yung20a}. $\dot N_{\rm ion}$ can be computed from \Halpha emission by 
\begin{equation}
\dot N_{\rm ion} = \frac{L({\rm H\alpha})}{1-f_{\rm esc}^{\rm LyC}} \times 7.35\times10^{11}\ {\rm erg^{-1}}, 
\end{equation} 
assuming Case B recombination at $T_{\rm e}=10^4\,\text{K}$ \citep{kennicutt_94, madau98}. We then obtain the production efficiency of ionising photons that do not escape from the galaxy, $\ksio$, assuming the Lyman continuum escape fraction ($f_{\rm esc}^{\rm LyC}$) is 0. Given the non-detection in the UV continuum, we are only able to derive a lower limit on $\log_{10}(\ksio/\text{[Hz/erg]})$ of $\gtrsim 26.4$ and $\gtrsim26.0$ for the A and B components, respectively.  

These $\ksio$ values are extreme even for very massive stars (or Pop~III stars): \citet{schaerer_2025} propose a maximum of $\log_{10}(\ksio/\text{[Hz/erg]}) \approx 25.8$, while for more extreme IMFs including only massive stars, the ionising efficiency of metal-free stars can reach up to $\log_{10}(\ksio/\text{[Hz/erg]}) = 26.1$ \citep{raiter_predicted_2010}. As previously noted by \citet{vanzella_extremely_2023} and \citet{nakajima_ultra-faint_2025}, the potential discrepancy could be due to the uncertainty on the sizes of the continuum- and emission-line-emitting regions, which would result in differential magnification for the emission lines and the UV continuum. Overall, the high $\ksio$ is a sign of a very young stellar population, with the potential presence of high-mass stars. 

\begin{figure}
\centering
\includegraphics[width=0.99\columnwidth]{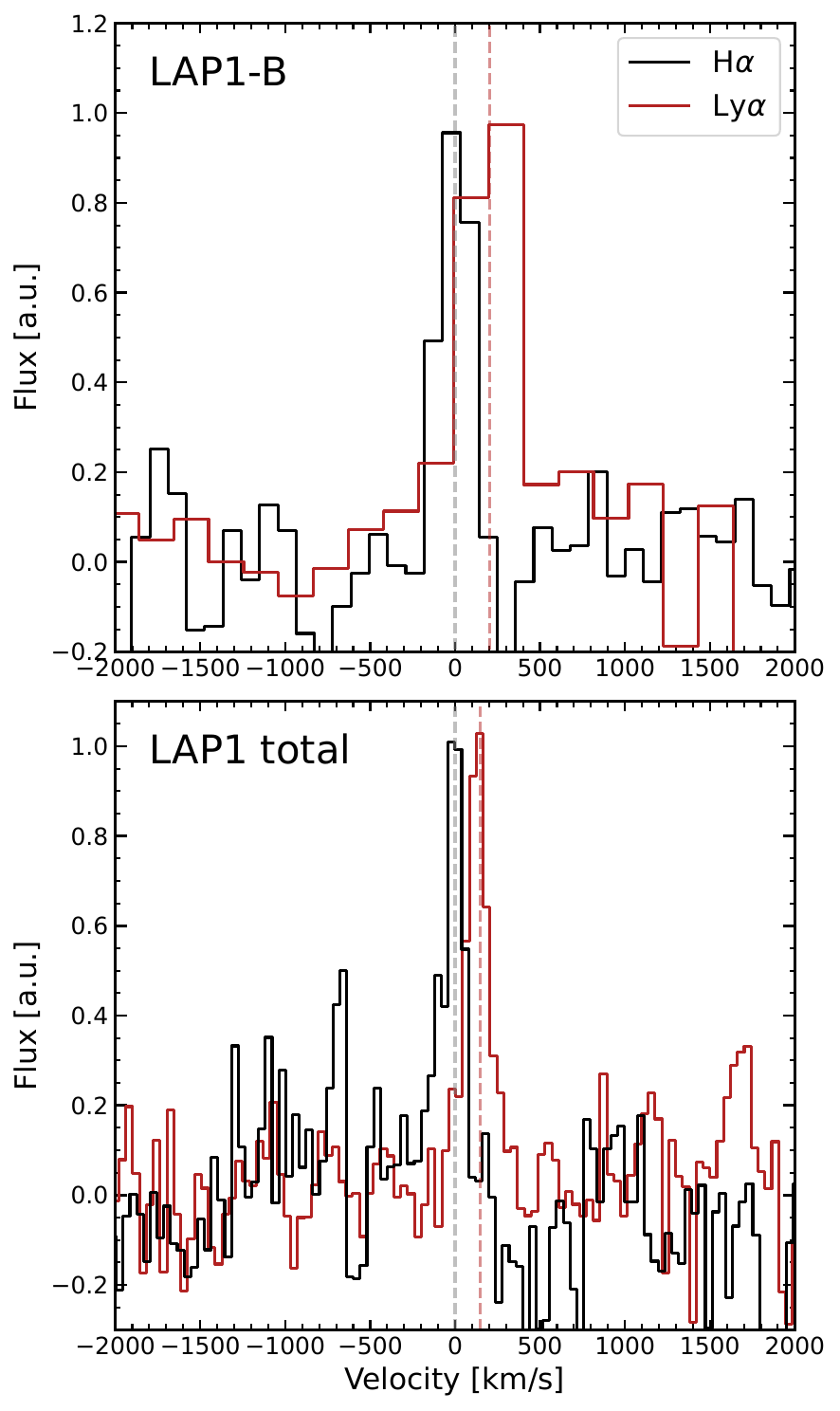}
\caption{Comparison of the \Lya and \Halpha profiles for LAP1-B from \MSA data (top panel) and the total system based on the \highresifs and VLT/MUSE data (bottom panel). The modest offsets of $150\text{--}200$~\kms\ indicate a modest neutral hydrogen column density.}
\label{fig.lya_offset}
\end{figure}

We show the comparison of the \Lya and \Halpha line profiles in Fig.~\ref{fig.lya_offset}. The high values of $f_{\rm esc}^{\Lya}$ are consistent with the low velocity offset between the \Halpha and \Lya emission lines, measured in the \MSA observations of the LAP1-B component as $130\pm30$~\kms, as both indicate a small amount of neutral gas along the line of sight (see Fig.~\ref{fig.lya_offset}). For LAP1-A, we do not have any medium- or high-resolution spectroscopy covering the \Lya emission, as PRISM-IFS observations do not have sufficient spectral resolution and VLT/MUSE observations do not have sufficient spatial resolution to separate LAP1-A and LAP1-B. Therefore, we use the VLT/MUSE observations to measure the velocity offset between \Lya and \Halpha in the total LAP1 integrated spectrum. \Lya is redshifted compared to \Halpha by 200$\pm$30 \kms. In the $\delta v$--\fescLya\ plane, LAP1 lies within the scatter of $z>4$ \Lya emitters (e.g., \citealt{Jones_lya_2025}), suggesting a low neutral hydrogen column density and/or a porous neutral-gas distribution around the star-forming regions.

The comparison of the \Lya and \Halpha line widths in both LAP1-B and the total system implies that there are no significant radiative transfer effects due to neutral gas or dust. Furthermore, the small velocity offset ($\lesssim200$~\kms) between the lines further supports the lack of neutral gas along the line of sight. Based on the modelling presented in \citet{Verhamme_2015}, a velocity offset of $<200$~\kms\ corresponds to a column density of N$_{\rm HI} \lesssim 10^{19.5}$ cm$^{-2}$. This low N$_{\rm HI}$ is consistent with stellar feedback ionising or clearing out the neutral gas surrounding the young star-forming regions. Given the low stellar age of the system (see \S\S~\ref{s.enrichment} and \ref{s.discussion}), this implies very rapid and efficient feedback.

\subsection{Metal enrichment}\label{s.enrichment}
%The observational constraints on LAP\,1 pertain exclusively to the ionised gas phase. No information is available on the neutral gas composition, though by analogy with metal-poor blue compact dwarf galaxies such as IZw\,18 and IZw\,36, the metal abundance of the neutral gas may be substantially lower than that of the ionised component. 
%
Given the low stellar mass of the LAP1 components and the low metallicity, we can estimate the oxygen mass in the ionised gas and the possible enrichment pathways. Following \citet{Chavez2014}, we can estimate the total ionised gas mass of the system as: %the total ionised gas mass is, 
\begin{equation}
    M_\mathrm{ion}
     \simeq 5 \times 10^{-34}
    \frac{L(\mathrm{H}\beta)\,m_\mathrm{p}}{\alpha_{\mathrm{H}\beta}^\mathrm{eff}
    \,h\nu_{\mathrm{H}\beta}\,n_\mathrm{e}}
    \simeq 6.8 \times 10^{-33}\,\frac{L(\mathrm{H}\beta)}{n_\mathrm{e}},
\end{equation}
where $M_\mathrm{ion}$ is given in \Msun, $L(\mathrm{H}\beta)$ is the observed \Hbeta luminosity in erg s$^{-1}$, $m_\mathrm{p}$ is the proton mass 
in g, $\alpha_{\mathrm{H}\beta}^\mathrm{eff}$ is the effective \Hbeta line recombination coefficient in cm$^3$ s$^{-1}$ for Case B in the low-density limit and $T = 10^{4}$~K, $h$ is the Planck constant in erg s, $\nu_\mathrm{H\beta}$ is the frequency corresponding to the \Hbeta 
transition in Hz, and $n_\mathrm{e}$ is the electron density in cm$^{-3}$.

This yields $M_\mathrm{ion} \sim 2 \times 10^4\,M_\odot$ for $n_\mathrm{e} = 10^2$\,cm$^{-3}$ and $M_\mathrm{ion} \sim 2 \times 10^3\,M_\odot$ for $n_\mathrm{e} = 10^3$\,cm$^{-3}$. The measured oxygen abundance of $12 + \log(\mathrm{O/H}) = 6.5$ corresponds to a mass ratio $(\mathrm{O/H})_\mathrm{mass} = 5.06 \times 10^{-5}$, implying a total oxygen mass in the ionised gas of $M_\mathrm{O} \sim 1\,M_\odot$ in LAP1-B for the lower-density case, and $M_\mathrm{O} \sim 0.1\,M_\odot$ for the higher-density case.

According to the chemical yield models of \citet{Molla_2012}, a $10^6\,M_\odot$ cluster ejects a total oxygen mass of $\sim 3050\,M_\odot$ over the supernova (SN) active phase. Since SNe do not contribute to the cluster ejecta until $\sim 2.0$\,Myr, the effective duration of the SN phase over the computed 20\,Myr baseline is 18.0\,Myr, giving a mean oxygen ejection rate of $\sim 170\,M_\odot\,\mathrm{Myr}^{-1}$ per $10^6\,M_\odot$, with only weak dependence on initial metallicity. Scaled to the ionising cluster mass of LAP1 ($M_\ast \sim 10^4\,M_\odot$), the average oxygen production rate is $\sim 1.7\,M_\odot\,\mathrm{Myr}^{-1}$.

However, the instantaneous oxygen production rate declines exponentially with time, being approximately an order of magnitude higher at $\sim 5$\,Myr than at 20\,Myr \citep{Molla_2012}. Consequently, most of the oxygen enrichment is concentrated within the first few Myr of SN activity. During the first Myr alone, a $10^4\,M_\odot$ cluster is expected to produce $\sim 5\,M_\odot$ of oxygen, well in excess of the $0.1$--$1\,M_\odot$ inferred for LAP1. The observed oxygen abundance can therefore be accounted for by the very first SN events, or at most a handful of them. We note that for cluster masses below $\sim 10^4\,M_\odot$, stochastic sampling of the IMF becomes significant \citep{Cervino2004, Cervino2006}, introducing substantial uncertainty in all derived quantities \citep{Rossi_26}.

Our results are consistent with more detailed modelling of self-polluted Pop~III clusters by \citet{rusta_metal-polluted_2025}, which combines chemical enrichment models with photoionisation models to predict the \OIIIL[5008]/\Hbeta ratio as a function of metal pollution within the Pop~III cluster. Based on their simulations, such self-pollution can be achieved in $\sim3$~Myr, similar to our results. 

\subsection{Connection to the star clusters at high-z}\label{s.discussion}

In this section, we discuss the implications of the confirmed metal-poor, low stellar mass nature of LAP1 and its large apparent dynamical-to-stellar mass ratio (albeit based on upper limits), and we attempt to identify the physical nature of this system. In recent years, \jwst has uncovered a wealth of low-metallicity galaxies at high redshift \citep[][]{morishita_enhanced_2024, vanzella_extreme_2024, Fujimoto_2025, cullen_jwst_2025, Vanzella_lap2_2026}, some with very low estimated metallicities and stellar masses below $10^{7}\,\Msun$. Many of these systems are magnified by factors of $>30$, often lying close to the critical line of the foreground lensing cluster. Furthermore, many are spatially unresolved or only marginally resolved, with sizes inferred from NIRCam imaging of $\lesssim10$ to tens of pc, comparable to LAP1. Their stellar masses fall in the range $10^{4}\text{--}10^{6}\,\Msun$. 

\begin{figure}
\centering
\includegraphics[width=0.99\columnwidth]{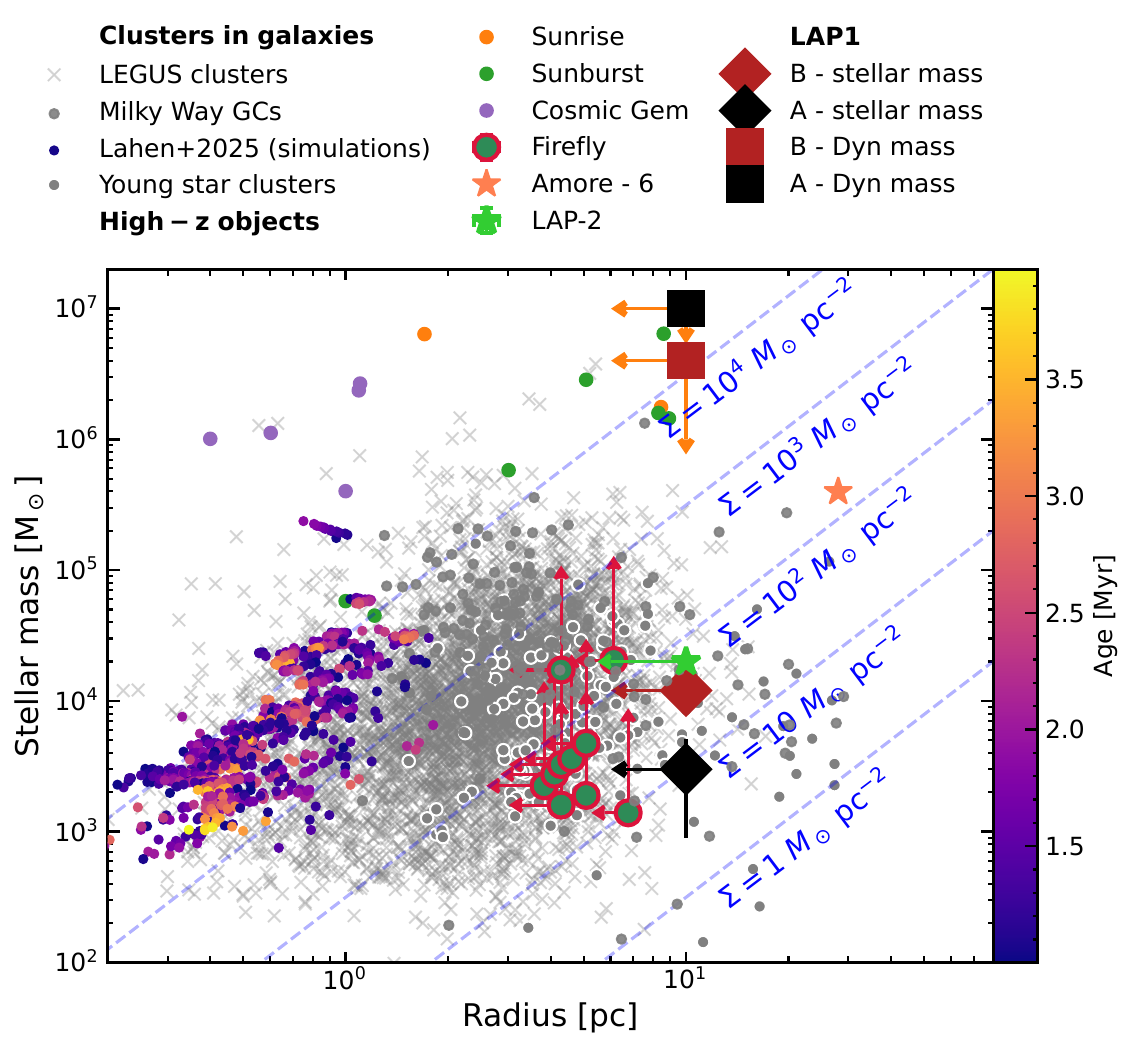}
\caption{Stellar (from the \Hbeta luminosity) and dynamical masses versus sizes for the LAP1 components, compared with those of the Firefly Sparkle \citep[][]{Mowla_2024}, Sunburst arc \citep[][]{rivera_thorsen_2017}, Sunrise arc \citep[][]{Vanzella_sun_2022, Vanzella_sun_2023} and Cosmic Gems \citep[][]{Adamo_2024}. We also include the LAP-2 \citep{Vanzella_lap2_2026} and AMORE6 objects \citep[][]{Morishita_2025}. Milky Way globular clusters \citep[][]{Gieles_2011} and young star clusters in star-forming spiral galaxies in the Local Volume (distances $<16$~Mpc; \citealt{Brown_2021}) are also shown for reference. We compare the observational results to the simulations of young clusters from \citet{Lahen_2025}.}
\label{fig.cluster_density}
\end{figure}

In Fig.\,\ref{fig.cluster_density}, we compare the stellar mass of LAP1 with those of other low-metallicity, low-mass high-redshift objects and star clusters detected in high-redshift galaxies. We also include the dynamical masses for LAP1 components. The sizes and stellar masses of LAP1-A and LAP1-B are consistent with those of local young star clusters and clusters observed in high-redshift galaxies (e.g. the Firefly Sparkle, Cosmic Gems, Starburst arc and Sunburst arc), suggesting that these primitive objects are possibly the progenitors of star clusters within their host galaxies or, for the more massive systems, already constitute young, low-metallicity star clusters. Furthermore, our measurements are consistent with simulations of young star cluster formation in low-metallicity environments from \citet{Lahen_2025}. 

The fact that we detect only the bare, compact source near the critical line, with no extended emission, indicates that any diffuse host galaxy component has a surface brightness too low to be detected even with \jwst/NIRCam. What makes the LAP1 system particularly remarkable, however, is the detection of two distinct sub-components separated by $\lesssim130$~pc in projection in the source plane \citep{vanzella_extremely_2023}.

We note that LAP1-A lies very close to the lens caustic, which introduces significant uncertainty in both the magnification and the fraction of the source flux captured within the image. The magnification is estimated to span $500\text{--}20{,}000$, and throughout this work we adopt a conservative lower limit of 500. Given its proximity to the caustic, it is also possible that the lens is preferentially magnifying only a portion of the total source, rendering both the inferred size and total flux uncertain. It is therefore possible that we are observing only a portion of a star cluster with a stellar mass of $10^{2\text{--}3}\,\Msun$, similar to the recent constraint on Earendel \citep{Pascale_2025}. 

The most striking feature that distinguishes LAP1 from other known low-metallicity systems is the non-detection of the broadband continuum in both rest-frame UV and optical imaging. This strongly suggests that the system is in the very earliest stage of its first star formation episode, with the stellar population not yet old enough to produce a detectable continuum, similar to another Pop~III cluster candidate, Hebe \citep[][]{Maiolino_2026, Uebler_2026}. This interpretation is further supported by the observed \Lya emission, which demonstrates that the system is essentially unobscured by dust -- ruling out dust attenuation as the cause of the UV continuum non-detection, since any appreciable dust column would also suppress the \Lya emission.

The star formation rate estimated from the UV is typically calibrated over a 100\,Myr timescale \citep{kennicutt_star_2012}. Renormalising to a 10\,Myr timescale -- consistent with the \Halpha calibration -- the UV-derived SFR of $<0.01\,\Msun\,\text{yr}^{-1}$ is fully consistent with the \Halpha-derived value of $\sim0.006\,\Msun\,\text{yr}^{-1}$, and both are consistent with those of ultra-faint dwarfs (UFDs; see e.g. \citealt{Salvadori_2014}). The agreement between these two tracers, together with the UV continuum non-detection, is consistent with a single, recent burst of star formation with no prior episodes; had earlier bursts occurred, they would have produced stars with detectable UV and optical continuum emission. A lower limit on the age of LAP1 can be placed from the already measurable metal enrichment of the nebular gas. Simulations of Population~III star formation suggest that measurable metal pollution requires at least $\sim3$~Myr of evolution for high-mass ($>100$\,\Msun) Pop~III stars \citep{rusta_metal-polluted_2025}. An upper limit on the age is set by the high \Halpha EW ($>1200$\,\AA) and non-detection of the UV continuum. Based on the Starburst99 models, this would imply ages of $<2\text{--}4$\,Myr. Combining this with the lower limit on the age based on the metal pollution in the object (see \S~\ref{s.enrichment}), we conclude that the stellar age of LAP1 is approximately 2--5\,Myr.

This picture is consistent with the growing body of evidence from deep NIRCam imaging of lensed cluster fields, which reveals that a large fraction of high-redshift galaxies host individual compact star-forming regions that can account for up to $\sim50\%$ of the total stellar mass of the underlying galaxy \citep[e.g.][]{Bradac_2024, vanzella_pristine_2025, Fujimoto_2025}. In this regime, the more diffuse underlying host galaxy falls below the detection threshold in both continuum and emission lines, leaving only the compact, high-surface-brightness clusters visible, similarly to what we observe for the LAP1 system. However, in the case of LAP1, we are probably witnessing the formation of the compact clusters in the galaxy, rather than observing the already formed clusters as in other systems \citep[e.g.][]{Adamo_2024, Bradac_2025}.

Many of the properties (formation age, SFR, upper limit on the dynamical mass) of LAP1-A and LAP1-B are also consistent with those of ultra-faint dwarfs, as already pointed out by \citet{nakajima_ultra-faint_2025}. Both high-z clusters and UFD progenitors have low metallicities ($<$5\%; \citealt{Tolstoy_09}), young stellar ages and low SFRs. Therefore, it is challenging to differentiate between the two scenarios. In the case of the LAP1 system, the picture is more challenging as there are two sources within $\lesssim130$~pc \citep[][]{vanzella_extremely_2023}. Although a pair of UFDs is known \citep[e.g.][]{McQuinn_2024}, their separation is over 30~kpc at $z=0$. However, the stellar-cluster and UFD scenarios can be distinguished by either constraining the \Mdyn or detecting any extended nebular emission between the A and B components. If the measured \Mdyn using high spectral and spatial resolution observations (such as with future E-ELT IFS data) is log(\Mdyn)$\gtrsim$6, this would indicate a high dark matter fraction and hence confirm the UFD scenario. On the other hand, a detection of extended nebular (e.g. \Halpha) emission connecting the A and B components would confirm the presence of an underlying host galaxy, confirming the stellar cluster scenario.

\section{Conclusions}\label{s.conclusion}
In this work, we have presented a comprehensive analysis of the Lensed and Pristine 1 (LAP1) system, combining all available \jwst imaging and spectroscopic observations, including new NIRSpec/IFS observations from the GA-NIFS survey in the G395H grating alongside PRISM/CLEAR IFS and MSA/R1000 data. Together, these observations cover the rest-frame UV through optical emission of this remarkable system. Our main conclusions are as follows:

\begin{enumerate}
\item We detect \OIIIL[5008], \Halpha, and \Hbeta emission in both the LAP1-A and LAP1-B components across all three spectroscopic configurations. Using the metallicity calibrations of \citet{Isobe_2026}, whose low-metallicity regime is calibrated on stacked spectra of JWST-selected extremely metal-poor galaxies at high-z, we measure oxygen abundances of $12+\log(\mathrm{O/H}) = 6.8 \pm 0.2$ and $6.5 \pm 0.2$ for the A and B components, respectively. These values are  $\sim0.2$\,dex higher than those reported in previous analyses of this system \citep{vanzella_extremely_2023, nakajima_ultra-faint_2025}, which we attribute to our choice of empirical calibrations derived from \jwst observations rather than high-ionisation photoionisation models. LAP1 remains one of the most metal-poor star-forming systems currently known.

\item Using NIRCam imaging from the CANUCS and PEARLS surveys, we created a stack of all available filters, as well as separate stacks of the filters covering the rest-frame UV and rest-frame optical emission. We do not detect continuum emission from any component of LAP1 in either the rest-frame UV or optical. These non-detections place $3\sigma$ upper limits on the stellar masses of $\log_{10}(M_\star/\Msun) < 3.0$ and $< 3.7$ for LAP1-A and LAP1-B, respectively. The combination of an undetected continuum with bright nebular emission lines implies that LAP1 is caught at the onset of its very first episode of star formation, with no significant prior build-up of stellar mass.

\item We do not robustly detect \CIVall emission in the NIRSpec-MSA R1000 data of LAP1-B1. Both direct spectral fitting and an independent bootstrapping analysis over 5000 realisations of the individual exposures yield a significance of only $\sim2.1\text{--}2.6\sigma$. We verify this by using an independent reduction from the Dawn \jwst Archive. As a consequence, we find no strong evidence for a hard ionising continuum consistent with Population~III stellar populations or direct collapse black holes. The resulting upper limit on $\log_{10}(\mathrm{C/O}) < -0.06$ is consistent with both Pop~II and Pop~III chemical enrichment channels. However, the tentative \CIVall detection is encouraging and prompts deeper observations for confirmation.

\item From the spectrally resolved \Halpha emission in the \highresifs data, we measure intrinsic line widths of $180^{+37}_{-35}$~\kms\ and $64^{+22}_{-26}$~\kms\ for the A and B components, respectively. We find no evidence for a physical velocity gradient in either component; apparent gradients are attributed to a systematic effect in the NIRSpec-IFS instrument model. Assuming virialised gas and adopting \Halpha-based effective radii of $<$10\,pc for A and B, we estimate dynamical masses of $\log_{10}(M_{\rm dyn}/\Msun) < 7.6$ and $<6.6$, respectively. However, we note that these are conservative upper limits given the uncertainty on the size of the \Halpha emitting regions. Adopting a cluster size of $<1$~pc, as predicted by theoretical simulations of low-metallicity clusters, would lower the dynamical mass by 1~dex.

\item We find a small velocity offset of \Lya relative to \Halpha of $200\pm20$ and $130\pm30$~\kms\ for the total LAP1 system and LAP1-B, respectively. Together with the high estimated escape fractions ($f_{\rm esc}^{\Lya}$ of $0.8\pm0.3$ and $0.5\pm0.1$), this indicates a low neutral gas column density surrounding the stellar clusters.

\item The combination of extremely low stellar mass, small physical size, and high surface mass density places both components of LAP1 in the same region of parameter space as young massive star clusters observed in gravitationally lensed high-redshift galaxies (e.g., Firefly Sparkle, Cosmic Gems, Sunburst arc). We interpret LAP1 as likely harbouring the earliest stages of star cluster formation within a low surface brightness host galaxy too diffuse to be detected in the current data. This interpretation is consistent with the non-detection of the UV continuum and the very high equivalent widths of \Halpha and \Lya.
\end{enumerate}

Nevertheless, the nature of the ionising sources in this system remains ambiguous, and confirmation of Population~III stellar populations will require deeper spectroscopy capable of robustly constraining \CIVall or \HeIIL[1640] emission.

\section*{Acknowledgements}

We would like to thank Maruša Bradač, Max Pettini and Angela Adamo for fruitful discussions that helped us interpret our results. We would also like to thank Natalia Lahén for providing results from her simulations. This work is based on observations made with the NASA/ESA/CSA James Webb Space Telescope. The data were obtained from the Mikulski Archive for Space Telescopes at the Space Telescope Science Institute, which is operated by the Association of Universities for Research in Astronomy, Inc., under NASA contract NAS 5-03127 for JWST. These observations are associated with program 4528, 1908 and 4750..
JS, RM, FDE, XJ and GCJ acknowledge support by the Science and Technology Facilities Council (STFC), ERC Advanced Grant 695671 ``QUENCH'' and the UKRI Frontier Research grant RISEandFALL. RM also acknowledges funding from a research professorship from the Royal Society.
EB acknowledges the support of the ``Ricerca Fondamentale 2024'' INAF program (GO grant ``A JWST/MIRI MIRACLE: Mid-IR Activity of Circumnuclear Line Emission'' and RSN1 mini-grant 1.05.24.07.01).
SC acknowledges support from the European Union (ERC, WINGS, 101040227).
SA acknowledges grant PID2021-127718NB-I00 funded by the Spanish Ministry of Science and Innovation/State Agency of Research (MICIN/AEI/10.13039/501100011033).
H\"U acknowledges support by the Max Planck Society through the Lise Meitner Excellence Program. H\"U acknowledges funding by the European Union (ERC APEX, 101164796). Views and opinions expressed are however those of the authors only and do not necessarily reflect those of the European Union or the European Research Council Executive Agency. Neither the European Union nor the granting authority can be held responsible for them.
PB acknowledges financial support from the Italian Space Agency (ASI)
through contract ``Euclid - Phase E'' and from the INAF Grants ``The
Big-Data era of cluster lensing'' and ``Probing Dark Matter and Galaxy
Formation in Galaxy Clusters through Strong Gravitational Lensing''.
This project received funding from the ERC Starting grant
NEFERTITI H2020/804240 (PI: Salvadori)
MP acknowledges support through the grants PID2021-127718NB-I00, PID2024-159902NA-I00, and RYC2023-044853-I, funded by the Spanish Ministry of Science and Innovation/State Agency of Research MCIN/AEI/10.13039/501100011033 and El Fondo Social Europeo Plus FSE+.
EV and MM acknowledge financial support through INAF GO Grant 2024 ``Mapping Star Cluster Feedback in a Galaxy 450 Myr after the Big Bang'', the project PRORIS - COSMOWEB ``A new era for cosmology:  exploiting the JWST revolution'', and the INAF ``Ricerca Fondamentale 2024'' grant ``Probing Dark Matter and Galaxy Formation in Galaxy Clusters through Strong Gravitational Lensing''. 

%%%%%%%%%%%%%%%%%%%%%%%%%%%%%%%%%%%%%%%%%%%%%%%%%%
\section*{Data Availability}

The datasets were derived from sources in the public domain: JWST/NIRSpec MSA and JWST/NIRCam data from the MAST portal: \url{https://mast.stsci.edu/portal/Mashup/Clients/Mast/Portal.html}.



\bibliographystyle{mnras}
\bibliography{mybib, mybib_add} % if your bibtex file is called example.bib

%%%%%%%%%%%%%%%%%%%%%%%%%%%%%%%%%%%%%%%%%%%%%%%%%%

%%%%%%%%%%%%%%%%% APPENDICES %%%%%%%%%%%%%%%%%%%%%
\appendix

\section{Kinematical models}\label{s.app.kin}

\begin{figure}
    \centering
    \includegraphics[width=0.9\columnwidth]{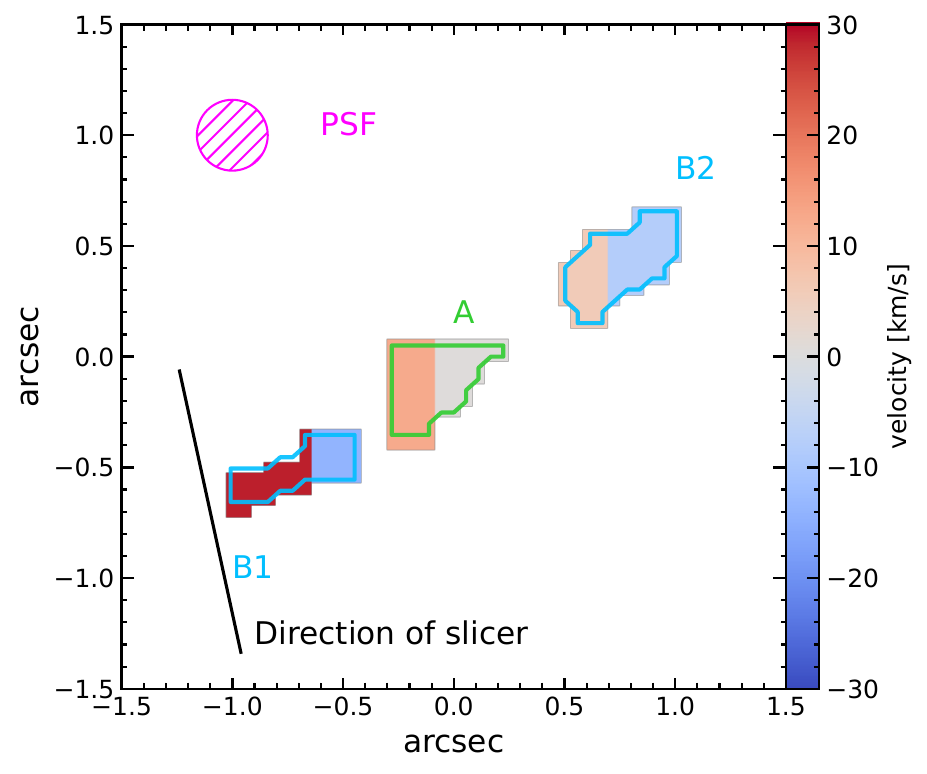}
    \caption{Velocity map of LAP1. We highlight the location of each of the components, along with the direction of the slicer as a black line. The magenta hatched circle shows the size of the PSF at 5~$\mu$m (observed wavelength of \Halpha). The velocity gradient is perpendicular to the direction of the slicer, which is a well-known issue with the NIRSpec-IFS model.  }
    \label{fig.vel_grad}
\end{figure}

We perform the kinematical modelling by splitting each component (A, B1 and B2) into two regions along the direction in which the objects are stretched by the lensing magnification. We do not have sufficient SNR to perform spaxel-by-spaxel fitting of the detected components; hence, we fit spatially binned spectra. Given the limited SNR of the spectra, we fitted only the \Halpha emission line, as it is the best-detected line in our high spectral resolution observations. We show the velocity offset map in Fig.~\ref{fig.vel_grad}. 

We find a velocity offset of $32\pm21$~\kms, $12\pm23$~\kms\ and $5\pm20$~\kms\ for the B1, A and B2 components, respectively. We do not detect a significant velocity offset in any of the three detected components. Furthermore, the three velocity gradients are all aligned perpendicular to the direction of the slicer, which is a well-known issue with the NIRSpec-IFS model \citep[see e.g.][]{juodzbalis_direct_2025}. Given the large uncertainties and the alignment of the velocity gradient, we conclude that these are instrumental or noise effects and not a real velocity gradient.

%%%%%%%%%%%%%%%%%%%%%%%%%%%%%%%%%%%%%%%%%%%%%%%%%%

% Don't change these lines
\bsp	% typesetting comment
\label{lastpage}

\end{document}